\documentclass[a4paper,fleqn]{cas-dc}

\usepackage{booktabs} % For formal tables
\usepackage{setspace}
\usepackage[ruled]{algorithm2e} % For algorithms
\usepackage{epsf,amsmath,amssymb,amsfonts,times,bm,xcolor}
\usepackage{epsf,epsfig,graphicx}
\usepackage{enumerate}
\usepackage{srcltx}
\DeclareMathAlphabet{\mathitbf}{OML}{cmm}{b}{it}
\usepackage{psfrag}
\usepackage{multirow}
\usepackage{framed}
\usepackage{stfloats}% long eq
\usepackage{textcomp}
\usepackage{balance}
\usepackage{subcaption}
\usepackage{caption}
\usepackage{bbm}
\usepackage{slashbox}
\usepackage[numbers]{natbib}
\usepackage{soul}
\newtheorem{remark}{Remark}%[section]
\begin{document}
\sloppy
% Title portion
\let\WriteBookmarks\relax
\def\floatpagepagefraction{1}
\def\textpagefraction{.001}
\shorttitle{}
\shortauthors{C. Tran et~al.}

\title [mode = title]{An Improved Approach for Estimating Social POI Boundaries With Textual Attributes on Social Media}                      

\author[1,3]{Cong Tran}[orcid=0000-0001-7511-2910]

\address[1]{Department of Computer Science and Engineering, Dankook University, Yongin 16890, Republic of Korea}

\author[2]{Dung D. Vu}

\address[2]{Machine Learning R\&D, Korbit AI, Montreal QC H2T 2A3, Canada}

\author[4]{Won-Yong Shin}
\cormark[1]
\address[3]{Machine Intelligence \& Data Science Laboratory, Yonsei University, Seoul 03722}
\address[4]{School of Mathematics and Computing  (Computational Science and Engineering), Yonsei University, Seoul 03722}

\cortext[cor1]{Corresponding author. 
}

\begin{abstract}
It has been insufficiently explored how to perform density-based clustering by exploiting textual attributes on social media. In this paper, we aim at discovering a social point-of-interest (POI) boundary, formed as a convex polygon. More specifically, we present a new approach and algorithm, built upon our earlier work on \underline{so}cial POI \underline{b}oundary \underline{est}imation (\textsf{SoBEst}). This \textsf{SoBEst} approach takes into account both {\em relevant} and {\em irrelevant} records within a geographic area, where relevant records contain a POI name or its variations in their text field. Our study is motivated by the following empirical observation: a {\em fixed} representative coordinate of each POI that \textsf{SoBEst} basically assumes may be {\em far away} from the centroid of the estimated social POI boundary for certain POIs. Thus, using \textsf{SoBEst} in such cases may possibly result in unsatisfactory performance on the boundary estimation quality (BEQ), which is expressed as a function of the $\mathcal{F}$-measure. To solve this problem, we formulate a joint optimization problem of simultaneously finding the radius of a circle and the POI's representative coordinate $c$ by allowing to update $c$. Subsequently, we design an {\em iterative} \textsf{SoBEst}  (\textsf{I-SoBEst}) algorithm, which enables us to achieve a higher degree of BEQ for some POIs. The computational complexity of the proposed \textsf{I-SoBEst} algorithm is shown to scale {\em linearly} with the number of records. We demonstrate the superiority of our algorithm over competing clustering methods including the original \textsf{SoBEst}.
\end{abstract}

\begin{keywords}
Boundary estimation quality \sep Geo-tagged record \sep Representative coordinate \sep Social point-of-interest boundary \sep Spatio--textual data.
\end{keywords}

\maketitle

\section{Introduction}
\subsection{Background}
With the pervasiveness of Global Positioning System (GPS)-enabled mobile devices, services based on not only location-based social networks (LBSNs) such as Foursquare \cite{c04} and Flickr \cite{c03} but also online social networks with geo-tagging or check-in functionality such as Twitter \cite{c01,c21}, Facebook \cite{c02}, and Instagram \cite{c1i} have grown rapidly in recent years.
These services provide various platforms for hundreds of millions of users to share their location-tagged media content objects such as photos, videos, musics, and texts. For example, when users visit a point-of-interest (POI), they are likely to check in, upload photos of their visit with annotations, or post geo-tagged textual data via LBSNs \cite{c1lbsn,c3lbsn,c2lbsn,c1c} to describe their individual idea, feeling, or preference relevant to the POI.
As a result, many textual postings have been geo-tagged on social media. For example, there are currently more than 600 million tweets posted on Twitter per day,\footnote{www.internetlivestats.com.} and approximately $1\%$ of them are geo-tagged \cite{c1twg}, corresponding to 6 million geo-tagged tweets per day.

Given the availability of the location information from geo-tagged or check-in records, there has been a steady interest in studying a variety of research challenges on accurately identifying or recommending POIs using social media. In particular, the problem of identifying POIs has a broad range of applications including but not limited to geomarketing, event management, urban planning, and location-based services.
For example, POI identification can be used in {\em geomarketing}, which is one of the most rapidly evolving trends in marketing. As a marketing strategy of companies targeting promotion of  a shopping mall, flyers will be disseminated online  hopefully only to the people who come to visit the desired  POI. Thus, company managers will not only be aware of the explicit marketing zone but also reduce the marketing cost.

On the other hand, in general, an area-of-interest (AOI) \cite{c25,c15,c11,c10} has been considered as either a single connected region or a union of disconnected regions in which multiple neighboring geographic coordinates of POIs could be found. Thus, we can interpret the AOI as a special region which possesses significantly high popularity amongst the locals. While most studies have focused on identifying a POI \cite{c3,c2}, we focus on estimating an AOI due to the robustness in reflecting interests of many users. To illustrate this point, we would like to address that tourists' interest often involves many aspects of daily life, including sightseeing, dining, shopping, etc., which cannot be covered by a single POI. Instead, an AOI that may contain various POIs and their associated area (e.g., tourist attractions, popular venues, and restaurants) often within a walking distance of each other would be ideal to recommend to tourists.

\subsection{Motivation and Main Contributions}
\begin{figure}[t]
\centering {
\begin{subfigure}[]{0.24\textwidth}
        \includegraphics[width=\linewidth]{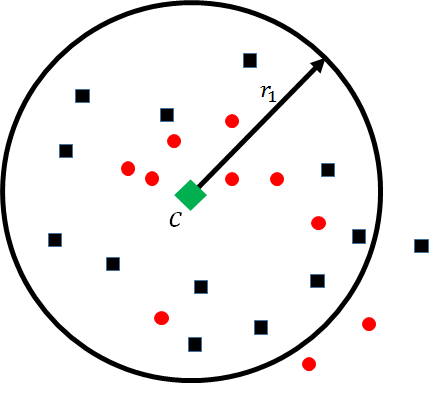}
        \caption{Circle 1}
        \label{fig:Circle1}
\end{subfigure}%
\begin{subfigure}[]{0.24\textwidth}
        \includegraphics[width=\linewidth]{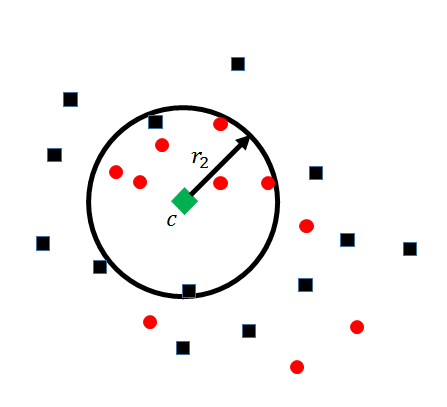}
        \caption{Circle 2}
        \label{fig:Circle2}
\end{subfigure}

}
\caption{A sample dataset, where $r_i$ for $i\in\{1,2\}$ and $c$ indicate the radius of a circle and the representative coordinate, respectively. Here, red circles and black squares indicate relevant and irrelevant records, respectively.}
\label{fig:1}
\end{figure}

In this paper, we would like to exploit {\em textual} attributes on social media in more accurately discovering {\em POI boundaries} corresponding to AOIs. To this end, by leveraging spatio--textual data on social media such as Twitter, we develop our own approach and algorithm, built upon a preliminary version \cite{cGeorich} of the current work for effectively \underline{est}imating a \underline{so}cial POI \underline{b}oundary, \textsf{SoBEst} (which was originally referred to as GeoSocialBound in \cite{cGeorich}). Our study differs from the most popular spatial clustering algorithm, so-called density-based spatial clustering of application with noise (DBSCAN)~\cite{c16}, in the sense that DBSCAN only exploits the set of geographic locations. For better understanding of our study, we start by revisiting both the notion of social POI boundaries and the \textsf{SoBEst} algorithm. A social POI boundary is defined as one cluster, which is given by a {\em convex polygon} geographically formed by taking the convex hull of geo-tagged records (e.g., geo-tagged tweets) that include a given textual description such as a POI name within a circle to be optimized.
In geomarketing, the marketing zone is expected to cover all geo-tagged records relevant to the desired POI with as little marketing cost  as possible; thus, it would be desirable to define the social POI boundary as a convex polygon that is the smallest region enclosing all relevant records in terms of cost efficiency as well as promotion effectiveness. Unlike most of existing spatial clustering algorithms, we are interested in taking into account both \textit{relevant} and \textit{irrelevant} records within a geographic area, where relevant records contain a POI name or its variations in their text field. Intuitively, we would expect the associated area to be neither too small nor too large. As two scenarios, a small POI boundary would cover only an insignificant fraction of relevant records while an unbounded region may contain many irrelevant records.
Given a POI name and its representative coordinate $c$, we aim to obtain a geographic area of interest (i.e., a social POI boundary) as large as possible, while maintaining a sufficiently high textual quality of the area, which is computed from portions of both relevant and irrelevant records. For example, suppose that there are two circles having the center $c$, indicated by a green diamond, and radii $r_1$ and $r_2$ in Fig.~\ref{fig:1}, where $r_1>r_2$. Red circles and black squares indicate relevant and irrelevant records within the circle $(c,r_i)$, respectively, for $i\in\{1,2\}$. In the figure, it is seen that Circle 1 has a worse annotation quality since it suffers from an overexpansion with a number of accompanying irrelevant geo-tags near the boundary of the circle. In this case, to reduce the number of irrelevant geo-tags, we need to decrease the radius of Circle 1, leading to an increment of the quality. To balance this trade-off, we define a {\em boundary estimation quality (BEQ)} as a function of the $\mathcal{F}$-measure \cite{c1L0}, denoted by $\mathcal{F}(c,r)$ for the radius $r$ of a circle to be found and the POI's representative coordinate $c$, where the $\mathcal{F}$-measure is a popular measure of a test's accuracy. From this measure, we expect to divide the plane into two separate regions (i.e., an interior and an exterior) by finding a sufficiently large circle containing a large number of relevant records and a small number of irrelevant records. In addition, we observe that the $\mathcal{F}$-measure tends to be hardly degraded up to a certain critical point with respect to the radius $r$ from the initially given POI's representative coordinate $c^{(0)}$. Based on this observation, an optimization problem in the sense of maximizing the BEQ including a coverage term $(\frac{r}{\bar{r}})^\alpha$, denoted by $\left(\frac{r}{\bar{r}}\right)^\alpha\mathcal{F}(c^{(0)},r)$, was formulated in \cite{cGeorich}, where $\alpha\ge0$ is the radius exponent, balancing between different levels of geographic coverage, and $\bar{r}$ is a normalization factor representing the maximum searching distance. To effectively solve this problem, the computationally efficient \textsf{SoBEst} algorithm was also presented in \cite{cGeorich}.

Although the approach in \cite{cGeorich} is the first attempt to perform spatial clustering using {\em textual attributes} on social media, we would like to raise the following concern: a {\em fixed} representative coordinate of each POI that \textsf{SoBEst} basically assumes may be {\em far apart} from the centroid of the estimated social POI for centain POIs. In consequence, using \textsf{SoBEst} in such cases may possibly result in unsatisfactory performance on the BEQ. We first empirically confirm that there are indeed many such POIs. Then, to overcome this problem, we introduce a joint optimization problem of simultaneously finding the radius $r$ of a circle and the POI's representative coordinate $c$. We also propose an {\em iterative} \textsf{SoBEst} algorithm, dubbed \textsf{I-SoBEst}, by iteratively updating $r$ and $c$ for improving the estimation performance.

We empirically evaluate the performance of our \textsf{I-SoBEst} algorithm in terms of $\left(\frac{r}{\bar{r}}\right)^\alpha \mathcal{F}(c,r)$ (i.e., BEQ) for various radius exponents $\alpha$ and POIs.
We verify the superiority of the proposed approach over two baselines and three state-of-the-art density-based clustering methods as well as the original \textsf{SoBEst} without updating $c$. In this paper, we first compare our algorithm to a variant of the original DBSCAN method, as a baseline, that does not inherently exploit the textual information on social media, where the original  DBSCAN in \cite{c16} is modified to make it suitable for performance evaluation under our problem setting. Note that, although there have been a variety of studies on designing state-of-the-art DBSCAN algorithms \cite{stdbscan, hdbscan, c15, c11, c10} in diverse applications, they do not reveal further improvement on the performance over our own variant of the original DBSCAN in our setting since all of them were not developed based on the textual attributes (i.e., relevant and irrelevant records).
Our experiments show that \textsf{I-SoBEst} significantly outperforms the modified DBSCAN by up to an order of magnitude.
We also compare our \textsf{I-SoBEst} algorithm to another baseline method \cite{ocsvm} and three state-of-the-art methods \cite{c23,newref4,lzpd} for discovering POI boundaries, and confirm the consistent superiority of \textsf{I-SoBEst} for various environments. Moreover, via performance evaluation, we identify the case where the gain of \textsf{I-SoBEst} over the original \textsf{SoBEst} without updating the POI's representative coordinate is promising. Our experiments exhibit that the BEQ of \textsf{I-SoBEst} further improves \textsf{SoBEst}'s by up to 16.75\% depending on experimental settings. Finally, we analytically and numerically show the computational complexity of \textsf{I-SoBEst}, which follows a {\em linear} scaling with the number of geo-tags.

\begin{figure*}[t]
\centering
\includegraphics[width=0.8\linewidth]{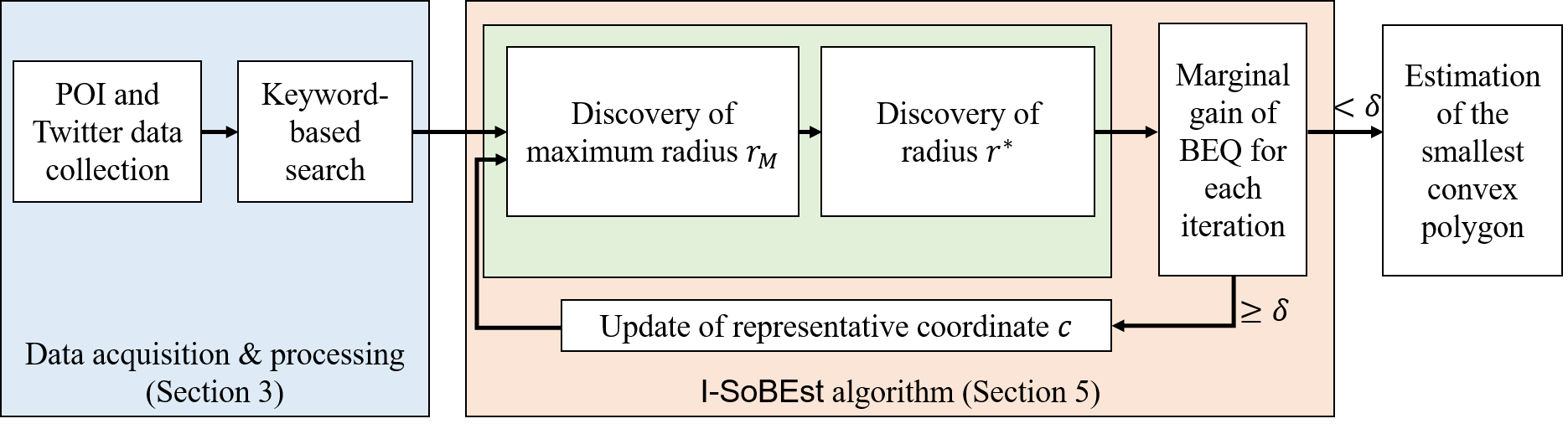}
\caption{The schematic overview of our \textsf{I-SoBEst} method.}
\label{fig:overall}
\end{figure*}

To the extent of our knowledge, this study subsuming our earlier work in \cite{cGeorich} is the first attempt to perform spatial clustering with high accuracy using {\em textual attributes} on social media.  The overall procedure of our \textsf{I-SoBEst} algorithm including data acquisition and preprocessing steps is illustrated in Fig.~\ref{fig:overall}, where notations and detailed descriptions for each block are shown in later sections.

\subsection{Organization and Notations}
The remainder of this paper is organized as follows.
In Section~\ref{sec:previouswork}, we review the related work.
Section~\ref{sec:dataacquisition} presents how to collect POIs and then how to search relevant tweets  for each POI.
In Section~\ref{sec:problemformulation}, we present our joint optimization problem for estimating social POI boundaries along with rigorous definitions.
Thereafter, we present the computationally efficient \textsf{SoBEst} algorithm and its enhanced \textsf{I-SoBEst} algorithm in Section~\ref{algorithms}; followed by  empirical evaluation and computational complexity analysis in Section~\ref{sec:computationalcomplexity}.
Finally, Section~\ref{sec:conclusion} summarizes the paper with some concluding remarks.

Table~\ref{tab:notation} summarizes the notations used throughout this paper. Some notations will be more precisely defined in the following sections when we introduce our problem formulation and algorithms.

\begin{table}[t!]
    \centering
    \caption{Summary of notations.}
    \label{tab:notation}

    \begin{tabular}{|l|p{6cm}|}
        \hline
        \multicolumn{1}{|c|}{\textbf{Notation}}&\multicolumn{1}{c|}{\textbf{Description}}                                                                                                \\ \hline
      $c$ & representative coordinate of a POI      \\ \hline
            $r$ & \begin{tabular}[c]{@{}l@{}} radius of a circle \end{tabular}     \\ \hline
            $\mathcal{F}(c,r)$ & \begin{tabular}[c]{@{}l@{}} $\mathcal{F}$-measure for a given circle $(c,r)$ \end{tabular}     \\ \hline
                        $r_\texttt{cover}$ &  approximate radius from a POI's representative coordinate covering the geographic area of each POI     \\ \hline 
            
                   $r^*$ & \begin{tabular}[c]{@{}l@{}} optimal radius of a circle \end{tabular}     \\ \hline
    $\alpha$   & \begin{tabular}[c]{@{}l@{}} radius exponent\end{tabular} \\ \hline
$\eta$      \begin{tabular}[c]{@{}l@{}}\end{tabular} & target quality threshold \\ \hline
    $\bar{r}$  \begin{tabular}[c]{@{}l@{}}\end{tabular} & searching distance  \\ \hline
                            \begin{tabular}[c]{@{}l@{}} $c^{(k)}$\end{tabular} & updated representative coordinate of a POI after the $k^\texttt{th}$ iteration \\ \hline
                \begin{tabular}[c]{@{}l@{}} $\delta$\end{tabular} & \begin{tabular}[c]{@{}l@{}} tolerance level of \textsf{I-SoBEst}\end{tabular} \\ \hline
                                  \begin{tabular}[c]{@{}l@{}}  $c^*$ \end{tabular}& finally updated representative coordinate of a POI     \\ \hline
 $\mathcal{D}(c,r)$ & set of relevant records within a circle $(c,r)$    \\ \hline
               $\mathcal{D}_\texttt{all}(c,r)$ & set of all geo-tagged records within a circle $(c,r)$   \\ \hline
$\Delta r$  \begin{tabular}[c]{@{}l@{}}\end{tabular} & sampling interval in $(0,\bar{r}]$  \\ \hline
    \end{tabular}

\end{table}

\section{Related Work}
\label{sec:previouswork}
Our proposed algorithm is related to the following  four broad fields of research: AOI discovery, density-based spatial clustering, spatio--textual similarity search, and clustering based on spatial and non-spatial attributes. \newline

{\bf AOI discovery.} The term AOI is also referred to as place-of-interest \cite{cMontoliu}, region-of-interest \cite{cTan}, or POI boundary \cite{c8,c23}. In a broad sense, the research on AOI discovery can be categorized into the following two problems: estimation of a small-scale AOI (i.e., a single POI boundary) and estimation of large-scale AOIs (i.e., multiple POI boundaries). 

First, the  approaches to discovering an AOI on a small scale were studied in \cite{c8,ocsvm,c23}, which lies in the same line as our work. In \cite{c8}, geo-tagged Flickr textual data, Wikipedia concepts, and links among them with a topical similarity method were employed to discover a POI boundary.   In \cite{ocsvm}, the boundary of an AOI was estimated using kernel density estimation and support vector machine (SVM) via geo-tagged Flickr photos. A two-phase algorithm for estimating a POI boundary with linear scaling complexity in the number of input records was proposed in \cite{c23}.

Second, previous studies on discovery of large-scale AOIs usually result in a list of areas containing multiple POIs \cite{c15,c11,cMontoliu,cTan,lzpd,newref5,newref6}. In \cite{c15,c11}, the authors presented a ranked list of attractive regions where many tourism spots are located for recommendation to travellers. Areas in which people tend to stay for a long time were discovered in \cite{cMontoliu}. In \cite{cTan}, both public regions and personal regions were extracted by using check-in histories on LBSNs. A technique that robustly identifies POIs' boundaries based on the local drastic changes of the data density was developed in \cite{lzpd}. In \cite{newref5}, an end-to-end framework for discovering POIs/AOIs from temporal properties and data attributes as well as geo-tagged photos was presented. In \cite{newref6}, AOIs were discovered by fusing both sentimental attributes of locations and multi-activity centers for each user.

{\bf Density-based spatial clustering.} DBSCAN \cite{c16} has been known as one of the most popular density-based  spatial clustering algorithms~\cite{c15,c11,c10} due to the capability of indexing separated clusters, the robustness in detecting outliers, and the ability to provide arbitrarily-shaped clusters. DBSCAN algorithms find multiple clusters based on any source of location information with reasonable computational complexity. As follow-up studies on DBSCAN, numerous sophisticated algorithms have been developed as follows: a generalized version of DBSCAN that differently measures the cardinality of neighborhoods of a point was proposed in \cite{gdbscan}; another DBSCAN algorithm that generates clusters based on both spatial and temporal attributes was presented in \cite{stdbscan}; and hierarchical DBSCAN was introduced in \cite{hdbscan} as an automated method that measures the stability of clusters in a clustering hierarchy to select proper input parameters of DBSCAN. Recently, several studies attempted to reduce the computational complexity of DBSCAN algorithms by either employing techniques such as bitmap indexing to support efficient neighbor grid queries \cite{dbscannew1} or transforming the problem of identifying a core point to a k-nearest neighbor (kNN) problem \cite{dbscannew2}.

On the other hand, density peaks clustering (DPC) achieving competitive performance in a non-iterative manner was recently developed in \cite{newref1,newref2,newref3,newref4} as an alternative to  DBSCAN as it is based on the assumption that cluster centers have higher density than their neighbors and are located sufficiently far from other points with higher local density.

{\bf Spatio--textual similarity search.} It is of importance to find spatially and textually closest objects to query objects. To offer compelling solutions to this problem, several algorithms \cite{cFelipe,cCong} were introduced. In particular, a method to answer queries containing a location and a set of keywords was presented in \cite{cFelipe}. An index framework for processing top-$k$ query that takes into account both spatial proximity and text relevancy was introduced in \cite{cCong}. Although these algorithms study the spatio--textual distance between objects, they are inherently different from our approach, which estimate a POI boundary utilizing the textually relevant and irrelevant records on social media such as Twitter.

{\bf Clustering based on spatial and non-spatial attributes.} There have been recent studies on the use of spatial and non-spatial attributes on social media to improve the clustering performance in various applications. In \cite{cGennip}, spectral clustering was applied to identify clusters among gang members based on both the observation of social interactions and the geographic locations of individuals. By utilizing the geo-tagged textual data to find correlations between textual descriptions (i.e., POI types) and locations, the problem of discoverying AOIs grouped by categories was also studied in \cite{c25}.

\section{Data Acquisition and Processing}
\label{sec:dataacquisition}
We describe how to collect POIs and Twitter data associated with their locations. Thereafter, for each POI, we elaborate on our approach to search for relevant tweets.

\subsection{Collecting POIs}
To obtain a set of POIs
and their representive coordinates, we use an open source database Geonames,\footnote{http://www.geonames.org.} which has a great amount of geographical information and concepts \cite{c09}. There are  various feature classes in the database such as spots, buildings, farms, roads, hills, rocks, etc., but the following four representative categorized types are taken into account along with four POIs: building (S.BLDG), stadium (S.STDM), museum (S.MUS), and park (L.PRK). The four POI types are summarized in Table~\ref{tab: five}, where POIs and their representative attributes in the table will be further described in the next subsection.

\begin{table}[t]
\centering
\caption{POI types and attributes of four POIs according to Geonames.}
\label{tab: five}
\resizebox{\columnwidth}{!}{%
\begin{tabular}{|p{2.8cm}|p{1.4cm}|p{3.6cm}|}
\hline
{\textbf{POI name}}    &\textbf{Geonames category}& \textbf{Representative coordinate} $c^{(0)}=(c^{(0)}_{\textrm{lat}},c^{(0)}_{\textrm{lon}})$  \\ 
\hline
\begin{tabular}[c]{@{}l@{}}{Empire State}\\{Building}\end{tabular}  &S.BLDG& ($40.74871^o, -73.98597^o$) \\ \hline
Dodger Stadium  & S.STDM& ($34.073611^o, -118.24^o$)   \\ \hline
Metropolitan Museum of Art & S.MUS&($40.77891^o,-73.96367^o$)  \\ \hline
Busch Gardens & L.PRK&($28.03363^o, -82.41648^o$)  \\ \hline
\end{tabular}
}
\end{table}

\subsection{Collecting Twitter Data}
We use Twitter Streaming Application Programming Interface (API) \cite{c1t,c07,c3t,c2t}, which has
been widely used to collect data for topic modeling \cite{c1t, c07}, network
analysis \cite{c2t}, and statistical content analysis  \cite{c3t}. Streaming API returns tweets matching a query provided by a Streaming API user. It was reliably found that although the Twitter Streaming API only returns at most a $1\%$ sample of all the tweets produced at a given moment, it returns an almost complete set of \textit{geo-tagged} tweets despite sampling \cite{c1twg}.

Our dataset consists of 5,992,223 geo-tagged tweets
 recorded from 643,506 Twitter users from July 29, 2015 to August 29, 2015 (about one month)  in the US. Note that this short-term (one month) dataset is sufficient to estimate social POI boundaries with high accuracy. We removed the content that was automatically created by other services such as Tweetbot, TweetDeck, Twimight, and so forth. We see that each tweet contains a number of entities that are distinguished by their attributed field names. For data analysis, we adopt the
following three essential fields from the data of tweets:
\begin{itemize}
    \item   \emph{text}: actual UTF-8 text of the status update;
    \item   \emph{lat}: latitude of the tweet's location;
    \item   \emph{lon}: longitude of the tweet's location.
\end{itemize}

 As shown in Table~\ref{tab: five}, four POIs placed in the US are used for our analysis, where a POI is selected for each Geonames category. Representative attributes of the four POIs are summarized in Table~\ref{tab: five}. The third column of Table~\ref{tab: five} represents the initial POI's representative coordinate $c^{(0)}=(c^{(0)}_{\textrm{lat}},c^{(0)}_{\textrm{lon}})$ provided by Geonames, where $c^{(0)}_{\textrm{lat}}$ and $c^{(0)}_{\textrm{lon}}$ denote the latitude and longitude measured in degrees, respectively.

\subsection{Searching Relevant Tweets}
Since users are able to tag a POI name  (e.g., \textit{\#buschgardens}) or insert it (e.g., \textit{Busch Gardens} and \textit{Busch Park}) in their tweets to describe their interest in a POI,
we can easily query all relevant tweets that contains the POI name.
On the other hand, a POI name may be misspelled or  have other words tacked on to it. Thus, we perform a keyword-based search by querying semantically coherent but different words for a POI.\footnote{Note that brown clustering can also be adopted to find semantically coherent variations of a POI, even if it is not taken into account in our study.} Relevant variations for an annotation (i.e., a POI name) would include the hashtag followed by the POI name, its abbreviated name, its synonyms (if any), etc. (refer to Table~\ref{tab:query}). Then, the dataset can be partitioned into two subsets of geo-tagged tweets with and without the annotated POI names (refer to Table~\ref{tab:input} in Section~\ref{sec:computationalcomplexity} for detailed statistics of these two subsets for each POI).
\begin{table}[t]
\centering
\caption{Search queries.}
\label{tab:query}

\begin{tabular}{|p{2.3cm}|p{5.2cm}|}
\hline
\multicolumn{1}{|c|}{\textbf{POI name}}                                        & \multicolumn{1}{c|}{\textbf{Search queries}}   \\ \hline
Empire State Building                                                 &\textit{\#empirestate, Empire State,  Empirestate}                                             \\ \hline
\begin{tabular}[c]{@{}l@{}}Dodger Stadium\end{tabular} & \textit{\#dodgerstadium, Dodger Stadium} \\ \hline
Metropolitan Museum of Art & \textit{\#themet, \#metropolitanmuseumofart, Metropolitan Museum of Art, The Met, Themet, Metropolitainmuseumofart}  \\ \hline

\begin{tabular}[c]{@{}l@{}}Busch Gardens\end{tabular} & \textit{\#buschgardens, Buschgardens, Busch Gardens, Busch Park}  \\ \hline

\end{tabular}

\end{table}

\section{Problem Description}
\label{sec:problemformulation}
In this section, we formulate our new optimization problem after formally defining some important terms used for our analysis.

\subsection{Definitions}

We start by introducing three important definitions in analyzing the estimation performance of social POI boundaries. As mentioned before, the relevance of the data to a POI varies according to the geographic distance between the POI's representative coordinate and the locations where the data are generated. The tweets posted in locations far away from the representative coordinate of a POI are unlikely to have textual description for the POI. Hence, it is remarkably important to formally define a social POI boundary and to estimate it with high accuracy. Unlike density-based clustering algorithms such as DBSCAN that find \textit{arbitrarily-shaped} multiple clusters, to characterize a social POI boundary, we focus only on finding the \textit{maximum possible distance} reachable from the POI's representative coordinate, thereby enabling us to significantly reduce the computational complexity compared to other density-based clustering approaches \cite{c15,c11,c10}.

\par \textit{Definition 1:} Given a circle that has the representative coordinate  $c$ of a POI and a certain radius $r > 0$, let $\mathcal{D}(c,r)$ denote the set of all geo-tagged records (e.g., geo-tagged tweets) that include a given textual description (i.e., a POI name) within the circle $(c,r)$. Then, a social POI boundary  is defined as a \textit{convex polygon} geographically formed by taking the convex hull of the given points in $\mathcal{D}(c,r)$.

\par As an example of geomarketing  services, note that it would be desirable to define the social POI boundary as a convex polygon that is the smallest region enclosing all relevant records due to the cost efficiency as well as promotion effectiveness. Contrary to LBSNs with check-in records and geo-tagged photos, geo-tagged tweets with (semantically) different textual descriptions from a POI name may occur especially on Twitter. Such irrelevant geo-tags may be acceptable since one cannot expect that all geo-tags in a given region contain the same textual description. For the sake of brevity, we denote geo-tags whose text contains and does not contain an associated POI name by \textit{``relevant"} and \textit{``irrelevant"} records, respectively. Similarly as in \cite{c26}, to quantitatively measure the number of relevant records, we introduce the quality of an annotation in a given area in the following definition.

\par \textit{Definition 2}: Given a textual description (i.e., a POI name), the quality of the description in a circle $(c,r)$ is defined as
\begin{align}
\label{eq:1}
{\textrm{Precision}{(c,r)}}& =\frac{|\mathcal{D}(c,r)|}{|\mathcal{D_\texttt{all}}(c,r)|}
 = \frac{TP{(c,r)}}{TP(c,r)+FP{(c,r)}},
\end{align}
which is the ratio of true positives to all predicted positives, where $c$ and $r$ denote the representative coordinate of a POI (corresponding to the center of the circle) and the radius of the circle, respectively. Here, $\mathcal{D}_{\texttt{all}}(c,r)$ denotes the set of all geo-tagged tweets  within the circle $(c,r)$ and is thus expressed as the sum of $TP{(c,r)}$ and $FP{(c,r)}$; $TP{(c,r)}=|\mathcal{D}(c,r)|$ is the number of relevant records inside the circle $(c,r)$; and $FP{(c,r)}=|\mathcal{D}_{\texttt{all}}(c,r)\setminus \mathcal{D}(c,r)|$ is the number of irrelevant records in the circle $(c,r)$.

\textbf{Example 1}: Let us recall the sample dataset having  two circles in Fig.~\ref{fig:1}. Then, red circles and black squares correspond to $TP(c,r_i)$ and $FP(c,r_i)$, respectively. The annotation quality of each circle is computed as
\begin{itemize}
\item  Circle 1: $\textrm{Precision}{(c,r_1)} =\frac{8}{20}=0.40$;
\item  Circle 2: $\textrm{Precision}{(c,r_2)} =\frac{6}{8}=0.75$.
\end{itemize}

\par In our work, as a measure of boundary estimation accuracy, we use the $\mathcal{F}$-measure to judge a boundary. 

    \par \textit{Definition 3}: Given a POI name and a circle $(c,r)$, the $\mathcal{F}$-measure is given by
    \begin{align}
    \label{eq:2}
&\mathcal{F}{(c,r)}=\frac{2\textrm{Precision}(c,r)\textrm{Recall}(c,r)}{\textrm{Precision}(c,r)+{\textrm{Recall}{(c,r)}}},
    \end{align}
which indicates the harmonic mean of $\textrm{Precision}(c,r)$ and $\textrm{Recall}{(c,r)}$. Here, \textrm{Recall}$(c,r)$ is the ratio of true positives to all actual positives, that is $\frac{TP{(c,r)}}{TP{(c,r)+FN{(c,r)}}}$; and $FN(c,r)$ is the number of relevant records outside the circle $(c,r)$.
\par It is shown that the value of $TP(c,r)+FN(c,r)$ is fixed for a given dataset. Thus, $\mathcal{F}(c,r)$ depends on $TP(c,r)$ and $FP(c,r)$ from (\ref{eq:1}). More precisely, the $\mathcal{F}$-measure increases with high $TP(c,r)$ and low $FP(c,r)$.

\textbf{Example 2:} For the two circles in Fig.~\ref{fig:1}, the $\mathcal{F}$-measure of each circle is computed as

\begin{itemize}
    \item Circle 1: $\mathcal{F}{(c,r_1)} =\frac{2\times 0.4 \times 0.8}{0.4+0.8}=0.53$;
    \item  Circle 2: $\mathcal{F}{(c,r_2)}  =\frac{2\times 0.75 \times 0.6}{0.75+0.6}=0.67$.
\end{itemize}

From Example 2, one can obviously see that $\mathcal{F}(c,r)$ becomes higher when there are relatively more relevant records and less irrelevant records inside the circle $(c,r)$.
\begin{figure*}[t]
\centering {
\begin{subfigure}[]{0.48\textwidth}
	\centering
        \includegraphics[width=0.75\linewidth]{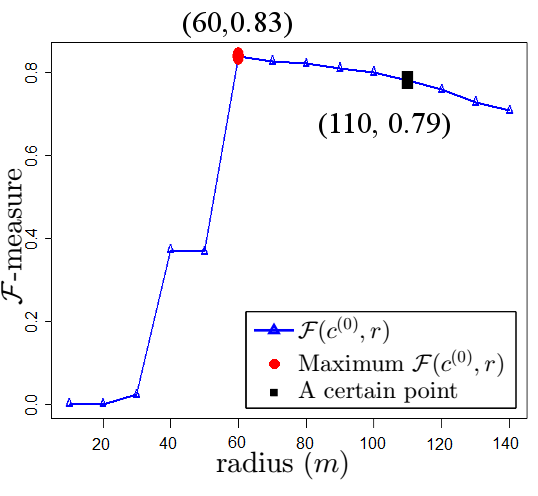}
        \caption{Empire State Building}
        \label{fig:Empirestatef1}
\end{subfigure}
\begin{subfigure}[]{0.48\textwidth}
	\centering
        \includegraphics[width=0.75\linewidth]{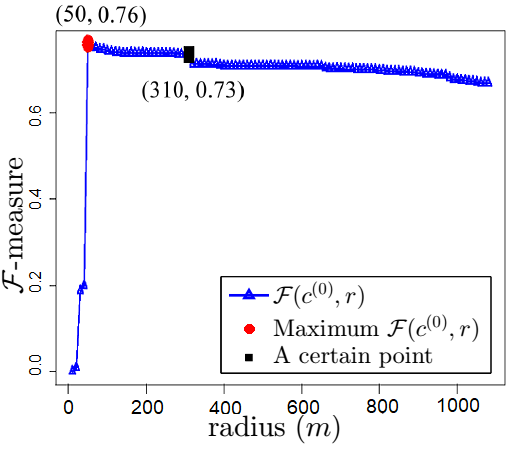}
        \caption{Dodger Stadium}
        \label{fig:Dodger11}
\end{subfigure}
\begin{subfigure}[]{0.48\textwidth}
	\centering
        \includegraphics[width=0.75\linewidth]{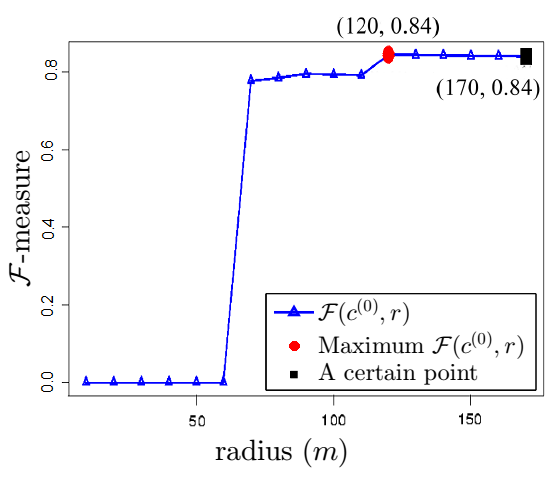}
        \caption{Metropolitan Museum of Art}        \label{fig:TheMet11 }
\end{subfigure}
\begin{subfigure}[]{0.48\textwidth}
	\centering
        \includegraphics[width=0.75\linewidth]{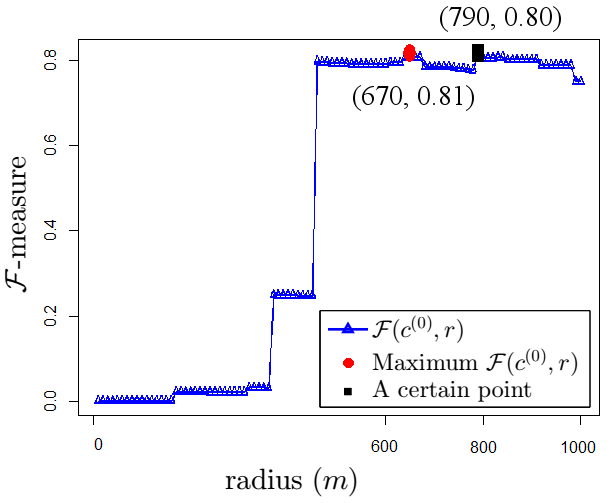}
        \caption{Busch Gardens}
       \label{fig:buschf1}
\end{subfigure}

}
\caption{The $\mathcal{F}$-measure according to the radius $r$.}
\label{fig: Fscore}
\end{figure*}

\subsection{Optimization Problem}
\label{sec:probfor}

In this section, we elaborate on our new optimization problem that aims at jointly finding the radius $r$ of a circle $(c,r)$ and the representative coordinate $c$. 

We begin by stating an interesting observation as follows. 
Fig.~\ref{fig: Fscore} depicts the value of $\mathcal{F}$-measure when varying the radius $r$ (in meters) from the initially given POI's representative coordinate $c^{(0)}$ for each POI. The maximum $\mathcal{F}(c^{(0)},r)$ for each POI is indicated by a red circle. Interestingly, it is observed that the  $\mathcal{F}$-measure tends to be hardly degraded up to a certain critical point with respect to the radius from the representative coordinate. For this reason, although it is desirable to provide a social POI boundary with the highest $\mathcal{F}$-measure, it would  also be good to considerably extend the social POI boundary (to cover more geographic region of the POI) at the cost of a \textit{slightly reduced value of} $\mathcal{F}(c^{(0)},r)$. One such application would include geomarketing as companies targeting promotion of a shopping mall may want to expand their marketing zone by disseminating flyers online to more people who are expected to come to visit the desired POI while slightly sacrificing the promotion accuracy (representing how many users accept the flyers as useful). For example, see the Empire State Building in Fig.~\ref{fig:Empirestatef1}---when the radius increases from 60m to 110m, $\mathcal{F}(c^{(0)},r)$ decreases from 0.837 (a red circle) to 0.792 (a black square), corresponding to only 5.68\% reduction in $\mathcal{F}(c^{(0)},r)$. Such a tendency can be found in the other examples.

In addition, we note that the approach in \cite{cGeorich} finds the optimal radius $r$ for the initially given representative coordinate $c^{(0)}$  of a POI provided by Geonames in the sense of maximizing the BEQ. However, motivated by the fact that a fixed representative coordinate of each POI may be far away from the corresponding centroid of all the relevant records in the estimated POI boundary for certain POIs, we introduce a {\em joint optimization} problem of finding both the radius $r$ of a circle $(c,r)$ and the POI's representative coordinate $c$ by allowing to update $c$. More specifically, we formulate a new problem that aims at jointly  finding both $r$ (in meters) and $c$ of a POI such that 
the corresponding BEQ is maximized, as follows:\footnote{As an alternative approach, the BEQ can be defined as the $\mathcal{F}_\beta$ measure for $\beta\gg1$, which is however not adopted in this work. Note that as $\beta>1$, more emphasis is imposed on recall than precision, which thus may lead to an extension of the social POI boundary compared to the case of $\beta=1$.}
        \begin{subequations} \label{eq:prob_ISoBEST}
                    \begin{align}
            &(c^*,r^*)=\arg\!\max_{c,r}\left(\frac{r}{\bar{r}}\right)^{\alpha} \mathcal{F}(c,r)\label{eq:4a}\\
            & \textrm{subject to}\quad 0< d(c,c^{(0)})+r \le \bar{r}\label{eq:4b}\\
            & \quad\quad\quad\quad\quad \textup{ Precision}(c,r)\ge \eta,\label{eq:4c}
            \end{align}
        \end{subequations}
where $\textup{Precision}(c,r)$ and $\eta \in (0,1)$ denote the textual quality in a circle $(c,r)$ in (\ref{eq:1}) and the target quality threshold (which will be specified in Section~\ref{sec:computationalcomplexity}), respectively. Here, $\alpha$ is the radius exponent; $d(c,c^{(0)})$ is the geographic distance between coordinates $c$ and $c^{(0)}$, where $c^{(0)}$ is the initial representative coordinate of a POI provided by Geonames; and $\bar{r}$ is the maximum searching distance and thus is set to an arbitrarily large value such that $\text{Precision}(c,\bar{r})\simeq 0$. In particular, the parameter $\alpha$ plays a crucial role in balancing different levels of coverage. For small $\alpha$, a social POI boundary with the almost highest $\mathcal{F}$-measure $\mathcal{F}(c,r)$ is found. As a special case, when $\alpha=0$, the original problem is boiled down to maximizing $\mathcal{F}(c,r)$ under two constraints (\ref{eq:4b}) and (\ref{eq:4c}). Equation (\ref{eq:4b}) is imposed so that the searching distance is constrained to the maximum allowable radius $\bar{r}$ from the representative coordinate $c^{(0)}$.  Equation (\ref{eq:4c}) is imposed so that the given minimum ratio of the number of relevant records in the circle $(c,r)$ to the number of all geo-tagged tweets in the same circle can be guaranteed.

\section{Proposed Estimation Method}\label{algorithms}

In this section, we describe our computationally efficient \textsf{I-SoBEst} algorithm that effectively solves the optimization problem in (\ref{eq:prob_ISoBEST}). For the completeness of this paper, we also review the original \textsf{SoBEst} algorithm in \cite{cGeorich}. 

In the algorithms, $\bar{r}$ is assumed to be arbitrarily large fulfilling $\textup{Precision}(c,\bar{r})\simeq 0$ for the representative coordinate $c$ of a POI. We suppose that two classified sets $\mathcal{D}(c,\bar{r})$ and $\mathcal{D}_\texttt{all}(c,\bar{r})$ are given as input, where $\mathcal{D}{(c, \bar{r})}$ and $\mathcal{D}_\texttt{all}(c,\bar{r})$ denote the sets of relevant records and all geo-tagged tweets, respectively, in the circle $(c, \bar{r})$. Since the geo-tags' coordinate is expressed in finite precision, we perform uniform sampling from $(0,\bar{r}]$ with a given sampling interval $\Delta r > 0$, where $\Delta r$ is set to a small value greater than the GPS error distance \cite{c1g}. Input parameters $\alpha$ and $\eta$ are determined according to the target levels of geographic coverage and the constraint of BEQ, respectively.
\subsection{\textsf{SoBEst} Algorithm}
\label{sec:BEst}

We summarize the overall procedure of our earlier work, \textsf{SoBEst}, with a slight modification in Algorithm~1. Since \textsf{SoBEst} is invoked multiple times within the \textsf{I-SoBEst} algorithm (refer to Section~\ref{sec:igeo} for more details), we hereby focus on describing the $k^{\texttt{th}}$ iteration of \textsf{SoBEst}, which basically consists of the following two phases.
\begin{itemize}
\item \textbf{Phase 1} (Lines 1--9 in Algorithm~1): We find the maximum radius $r_M$ such that $\textup{Precision}(c^{(k)},r)\ge \eta$ in (\ref{eq:4c}) for $0 < d(c^{(k)},c^{(0)})+ r \leq \bar{r}$ in  (\ref{eq:4b}).
\item \textbf{Phase 2} (Lines 11--18 in Algorithm~1): We find the optimal radius $r^{(k)}$ that maximizes  $(\frac{r}{\bar{r}})^{\alpha}\mathcal{F}(c^{(k)},r)$  under the constraints in (\ref{eq:4b}) and (\ref{eq:4c}).
\end{itemize}
Here, $c^{(k)}$ indicates the POI's representative coordinate after the $k^\texttt{th}$ iteration of \textsf{SoBEst}. In each phase, given a POI, we start by using a circle centered at $c$ with radius $r_{i} = \Delta r$. The radius is increased by $\Delta r$ for each step, and $r_M$ in Phase 1 or $r^{(k)}$ in Phase 2 is updated gradually if a certain condition is fulfilled. For each iterative step in Phase 1, the subroutine Filter($r_{i}, \mathcal{G}(c^{(k)},\bar{r})$) in Algorithm~2 is invoked to collect the set $\mathcal{G}(c^{(k)},r_{i})\subseteq \mathcal{G}(c^{(k)},\bar{r})$, where $\mathcal{G}(c^{(k)},r_{i})$ is either $\mathcal{D}(c^{(k)},r_{i})\textrm{ or } \mathcal{D}_\texttt{all}(c^{(k)},r_{i})$. To be specific, it follows that $\mathcal{G}(c^{(k)},r_{i}) = \{q_m|d(c^{(k)},q_m) \le r_{i}\}$, where $d(c^{(k)},q_m)$ denotes the geographic distance between $c^{(k)}$ and the $m^{\texttt{th}}$ geo-tag's coordinate $q_m$ in $\mathcal{G}(c^{(k)},\bar{r})$.\footnote{When the shortest path between two geo-tags' locations is measured along  the surface of the Earth, the distance between two locations can be computed according to the spherical law of cosines \cite{ccosine}.} Line 2 in Algorithm~2 can be implemented by using a searching algorithm such as \textit{R-trees} \cite{c1r}. The \textsf{SoBEst} algorithm finally returns $r^{(k)}, \mathcal{D}(c^{(k)},r^{(k)})$, and $\mathcal{F}(c^{(k)},r^{(k)})$ as the output of Algorithm 1.

%%%%%%%%%%%%%%%%%%%%%%%%%%%%%%%%%%%%%%%%%%%%%%%%%%%%%%%

\begin{table}[]

\renewcommand{\arraystretch}{1.1}
\centering
\begin{tabular}{l}
\hline
\textbf{Algorithm 1} \textsf{SoBEst} \cite{cGeorich}\\
\hline \textbf{Input}:$\mathcal{D}(c^{(k)},\bar{r}), \mathcal{D}_\texttt{all}(c^{(k)},\bar{r}), \Delta
r, \alpha, \eta, c^{(k)}$\\
\textbf{Output}: $r^{(k)}, \mathcal{D}(c^{(k)},r^{(k)}),\mathcal{F}(c^{(k)},r^{(k)})$\\
\textbf{Initialization}: $i\leftarrow1; r_i\leftarrow\Delta r; r_M\leftarrow0; r^{(k)}\leftarrow0;$\\ $N\leftarrow\frac{\bar{r}}{\Delta r}; TP(c^{(k)},r_i)\leftarrow0; FP(c^{(k)},r_i)\leftarrow 0; $\\ $ FN(c^{(k)},r_i)\leftarrow 0; \mathcal{F}{(c^{(k)},r_{i}) \leftarrow 0;} \mathcal{D}_\texttt{all}(c^{(k)},r_i)\leftarrow \emptyset;$ \\  $\mathcal{D}(c^{(k)},r_i)\leftarrow \emptyset; \mathcal{F}{(c^{(k)},r^{(k)})}\leftarrow 0;$ \\$\textup{Precision}(c^{(k)},r_i) \leftarrow 0; \textup{Recall}{(c^{(k)},r_i)}\leftarrow 0$\\

01: \hspace{0.3cm}\textbf{for} $i\leftarrow 1$  \textbf{to }$N$\\
02: \hspace{0.6cm}$\mathcal{D}_\texttt{all}(c^{(k)},r_{i})\leftarrow \textrm{Filter}(r_i,\mathcal{D}_\texttt{all}(c^{(k)},\bar{r}))$ \\\qquad \qquad\qquad\qquad\qquad\qquad(see Algorithm 2)\\
03:  \hspace{0.6cm}$\mathcal{D}(c^{(k)},r_{i})\leftarrow \textrm{Filter}(r_i,\mathcal{D}(c^{(k)},\bar{r}))$  \\\qquad\qquad \qquad\qquad\qquad\qquad(see Algorithm 2)\\
04:\hspace{0.6cm} $TP{(c^{(k)},r_j )}\leftarrow|\mathcal{D}(c^{(k)},r_j)|$\\
05:\hspace{0.6cm} $FP{(c^{(k)},r_j)}\leftarrow |\mathcal{D}_\texttt{all}(c^{(k)},r_j)\setminus \mathcal{D}(c^{(k)},r_j)|$\\
06:\hspace{0.6cm} $\textrm{Precision}(c^{(k)},r_{i}) \leftarrow\frac{TP{(c^{(k)},r_{i} )}}{TP{(c^{(k)},r_{i} )}+FP{(c^{(k)},r_{i} )}}$\\
07: \hspace{0.6cm}\textbf{if} $r_i>r_M$ \textbf{and} $\textrm{Precision}{(c^{(k)},r_i )}\ge\eta$ \textbf{then}\\
08:\hspace{1.2cm}$r_M\leftarrow r_i$\\
09: \hspace{0.6cm}$r_{i+1}\leftarrow r_{i}+\Delta r$\\
10: \hspace{0.3cm}$M\leftarrow \frac{r_M}{\Delta r}$\\
11: \hspace{0.3cm}\textbf{for} $i\leftarrow 1$  \textbf{to} $M$\\
12:\hspace{0.6cm} $FN{(c^{(k)},r_i )}\leftarrow|\mathcal{D}(c^{(k)},\bar{r})\setminus \mathcal{D}(c^{(k)},r_i)|$\\
13:\hspace{0.6cm} $\textrm{Recall}(c^{(k)},r_i) =\frac{TP{(c^{(k)},r_i )}}{TP{(c^{(k)},r_i )}+FN{(c^{(k)},r_i )}}$\\
14:\hspace{0.6cm} $\mathcal{F}{(c^{(k)},r_i)}=\frac{2\textrm{Precision}(c^{(k)},r_i)\textrm{Recall}(c^{(k)},r_i)}{\textrm{Precision}(c^{(k)},r_i)+{\textrm{Recall}{(c^{(k)},r_i)}}}$\\
15:\hspace{0.6cm} \textbf{if} $\left(\frac{r_i}{\bar{r}}\right)^{\alpha}\mathcal{F}{(c^{(k)},r_i)}\ge  \left(\frac{r^{(k)}}{\bar{r}}\right)^{\alpha}\mathcal{F}{(c^{(k)},r^{(k)})}$ \textbf{then}\\
16:\hspace{1.2cm}$r^{(k)}\leftarrow r_i $\\
17:\hspace{1.2cm}$ \mathcal{F}{(c^{(k)},r^{(k)})}\leftarrow \mathcal{F}{(c^{(k)},r_i)}$\\
18:\hspace{0.6cm} $r_{i+1}\leftarrow r_{i}+\Delta r$\\

19:\hspace{0.3cm}\textbf{return} $ r^{(k)}, \mathcal{D}(c^{(k)},r^{(k)}), \mathcal{F}{(c^{(k)},r^{(k)})}$

\end{tabular}
\end{table}

%%%%%%%%%%%%%%%%%%%%%%%%%%%%%%%%%%%%%%%%%%%%%%%%%%%%%%%%%%%%%
%%%%%

%%%%%%%%%%%%%%%%%%%%%%%%%%%%%%%%%%%%%%%%%%%%%%%%%%%%%%%%%%%

%%%%%%%%%%%%%%%%%%%%%%%%%%%%%%%%%%%%%%%%%%%%%%%%%%%%%%%%%%%%
\begin{table}[]
\renewcommand{\arraystretch}{1.1}
\centering
\begin{tabular}{l}
\hline
\textbf{Algorithm 2} Filter($r_i,\mathcal{G}(c^{(k)},\bar{r})$)\label{AL2}\\
\hline \textbf{Input}: $\mathcal{G}(c^{(k)},\bar{r})$ $(\mathcal{D}(c^{(k)},\bar{r}) \textrm{ or } \mathcal{D}_\texttt{all}(c^{(k)},\bar{r})), r_i$\\
\textbf{Output}: $\mathcal{G}(c^{(k)},r_i)$ for $r_i \in (0,\bar{r}]$\\
\textbf{Initialization}: $m \leftarrow 1; l \leftarrow |\mathcal{G}(c^{(k)},\bar{r})|$\\
01: \textbf{for} $m\leftarrow 1$  \textbf{to }$l$\\
02:  $\mathcal{G}(c^{(k)},r_i)\leftarrow \{q_m|d(c^{(k)},q_m) \leq r_i\}$\\

03: \textbf{return} $\mathcal{G}(c^{(k)},r_i)$

\end{tabular}
\label{tab:AL2}
\end{table}

Afterwards, by solving the convex-hull problem (e.g., \textit{quickhull} and an output-sensitive convex-hull algorithm \cite{c25a}), we can finally obtain the smallest convex polygon  that contains all the geo-tags in the set  $\mathcal{D}(c^{(k)},r^{(k)})$ quickly, corresponding to the estimated social POI boundary.

\subsection{\textsf{I-SoBEst} Algorithm}
\label{sec:igeo}

To effectively solve the joint optimization problem in (\ref{eq:prob_ISoBEST}), we introduce a low-complexity algorithm, named \textsf{I-SoBEst}. The \textsf{I-SoBEst} algorithm is designed in such a way that parameters $r$ and $c$ are \textit{iteratively updated} until a  given termination criterion is fulfilled.\footnote{In our study, we aim at offering a computationally efficient solution, rather than directly tackling the joint optimization problem along with the optimality.} The overall procedure is summarized in Algorithm~3. For each iterative step, the subroutine $\text{SoBEst}(\mathcal{D}(c^{(k)},\bar{r}),\mathcal{D}_\texttt{all}$ $(c^{(k)},\bar{r}), \Delta r, \alpha, \eta)$ is invoked to find the radius $r^{(k)}$ for a given POI's representative coordinate $c^{(k)}$, where $k$ indicates the iteration index and $r^{(k)}, \mathcal{D}(c^{(k)}, r^{(k)})$, and $\mathcal{F}(c^{(k)},r^{(k)})$ are returned as the output of \textsf{SoBEst}. In Line 4 of the algorithm, $p_l^{(k)}$ denotes the $l^\texttt{th}$ geo-tag's coordinate such that $p_l^{(k)}$ $\in$ $\mathcal{D}(c^{(k-1)},$ $r^{(k-1)})$, and thus $\frac{ \sum{p_l^{(k)}}}{L^{(k)}}$ represents the centroid of all relevant records in  the set $\mathcal{D}(c^{(k-1)},r^{(k-1)})$. The algorithm operates under the termination criterion in Line 6, where $\delta>0$ denotes the tolerance level determining the convergence speed and will be specified later. In  Section~\ref{sec:computationalcomplexity}, we shall show that the \textsf{I-SoBEst} algorithm outperforms the original \textsf{SoBEst} in terms of the BEQ $\left(\frac{r}{\bar{r}}\right)^\alpha \mathcal{F}(c,r)$ while maintaining the {\em linear} run-time complexity.

\begin{table}[t]
    \renewcommand{\arraystretch}{1.1}
    \centering
    \begin{tabular}{l}
        \hline
        \textbf{Algorithm 3} \textsf{I-SoBEst}
        \\
        \hline \textbf{Input}:$\mathcal{D}_\texttt{all}(c^{(0)},\bar{r}), \mathcal{D}(c^{(0)},\bar{r}), \Delta
        r , \alpha, \eta, \delta, c^{(0)}, \bar{r}$\\
        \textbf{Output}: $\mathcal{D}(c^*,r^*)$\\
        \textbf{Initialization}: $ k \leftarrow 0; r^* \leftarrow 0; c^* \leftarrow c^{(0)};   \mathcal{D}(c^*,r^*) \leftarrow \emptyset; $ \\$ (r^{(0)}, \mathcal{D}(c^{(0)}, r^{(0)}),  \mathcal{F}(c^{(0)},r^{(0)}))$ \\\qquad\qquad$\leftarrow  \textup{SoBEst}(\mathcal{D}(c^{(0)},\bar{r}), \mathcal{D}_\texttt{all}(c^{(0)}, \bar{r}),\Delta
        r, \alpha, \eta)$\\

        01:\hspace{0.3cm}\textbf{do}\\
        02:\hspace{0.6cm} $k \gets k+1 $ \\
        03:\hspace{0.6cm} $L^{(k)} = |\mathcal{D}(c^{(k-1)},r^{(k-1)})|$  \\
        04:\hspace{0.6cm} $c^{(k)} \gets \frac{ \sum{p_l^{(k)}}}{L^{(k)}} $\\

        05:\hspace{0.6cm} $(r^{(k)},\mathcal{D}(c^{(k)}, r^{(k)}),  \mathcal{F}(c^{(k)},r^{(k)}))$ \\\qquad\qquad\qquad$\leftarrow \textup{SoBEst}
        (\mathcal{D}(c^{(k)},\bar{r}), \mathcal{D}_\texttt{all}(c^{(k)}, \bar{r}),\Delta
        r, \alpha, \eta) $\\

        06:\hspace{0.3cm}\textbf{while} $\left(\frac{r^{(k)}}{\bar{r}}\right)^{\alpha}\mathcal{F}(c^{(k)},r^{(k)}) \!$
\\\qquad\qquad\qquad$-\left(\frac{r^{(k-1)}}{\bar{r}}\right)^{\alpha}\mathcal{F}(c^{(k-1)},r^{(k-1)}) \ge \delta$\\
        07:\hspace{0.3cm} $c^* \leftarrow c^{(k-1)}$\\
        08:\hspace{0.3cm} $r^* \leftarrow r^{(k-1)}$\\

        09:\hspace{0.3cm}\textbf{return} $\mathcal{D}(c^*,r^{*})$

    \end{tabular} \label{AL3}
\end{table}

\begin{remark}
Let us now state the convergence analysis of the \textsf{I-SoBEst} algorithm. Due to the fact that we have a finite number of geo-tagged records, there are a finite number of different representative coordinates $c^{(k)}$. It is obvious that the result of \textsf{SoBEst} is deterministic for given $c^{(k)}$, corresponding to a deterministic $\left(\frac{r^{(k)}}{\bar{r}}\right)^{\alpha}\mathcal{F}(c^{(k)},r^{(k)})$. From Line 6 of Algorithm 3, one can see that $\left(\frac{r^{(k)}}{\bar{r}}\right)^{\alpha}\mathcal{F}(c^{(k)},r^{(k)})$ is nonnegative and monotonically increasing after each iteration.  Hence, \textsf{I-SoBEst} converges to a local maximum in a finite number of iterations.
\end{remark}

\section{Performance Evaluation}
\label{sec:computationalcomplexity}
In this section, using the proposed \textsf{I-SoBEst} algorithm in Section~\ref{sec:igeo}, we first show experimental results for the BEQ and then analyze the overall average computational complexity. The Twitter dataset and the set of POIs in Section~\ref{sec:dataacquisition} are used for performance evaluation. Statistics of four POIs in Table~\ref{tab: five}, including $|\mathcal{D}(c^{(0)},\bar{r})|$ (i.e., the number of relevant records), $|\mathcal{D}_\texttt{all}(c^{(0)},\bar{r})|$, and $r_{cover}$, are shown in Table~\ref{tab:input}. As depicted in the table, since four POIs were selected according to each different Geonames category, both $|\mathcal{D}(c^{(0)},\bar{r})|$ and $|\mathcal{D}_\texttt{all}(c^{(0)},\bar{r})|$ are significantly different from each other.

\subsection{Experimental Setups}

For all POIs, we assume $\eta = 0.5$, which can be set to another value so as to control the constraint of BEQ, and $\Delta r = 10m$. Even if $\bar{r}$ is assumed to be arbitrarily large, we need to set $\bar{r}$ to a reasonable value in our experiment to efficiently reduce the complexity of the algorithm. In our study, we assume ${\bar{r}} = \gamma r_\texttt{cover}$, where $\gamma$ is a positive constant and $r_\texttt{cover}$ denotes  an \textit{approximate} radius from a POI's representative coordinate covering the geographic area of each POI and is obtained from Google Maps Geocoding API.\footnote{\text{https://developers.google.com/maps/documentation/geocoding/intro.}} The Google Maps Geocoding API returns an approximate bounding box of a given  POI, constructed by the southwest and northeast corners, and the value of $r_\texttt{cover}$ is obtained by computing the maximum of the two distances between the POI's representative coordinate and each of the southwest and northeast corners. While $\gamma$ can be properly determined according to the POI types (e.g., $\gamma = 1$ for a cafeteria and $\gamma = 10$ for an observation point), we use $\gamma = 10$ for all POIs for the sake of experimental simplicity. That is, we set $\bar{r}=10r_\texttt{cover}$ for a given radius $r_\texttt{cover}$ in Table~\ref{tab:input}, corresponding to the size of an approximate bounding box of each POI. It is then empirically demonstrated that the resulting $\text{Precision}(c,\bar{r})$ is negligibly small for all POIs, which coincides with our statement in Section~\ref{sec:probfor}. Moreover, we take into account the case where $\alpha \in [0, 1]$ since otherwise, the objective function in (\ref{eq:4a}) are largely dominated by the term $\left(\frac{r}{\bar{r}}\right)^\alpha$. More precisely, the value of $\alpha$ in geomarketing can be determined according to different marketing strategies. If a strategy aims to maximize the promotion accuracy (i.e., the $\mathcal{F}$-measure), then it is desirable to set $\alpha=0$, which results in the minimum marketing cost. If a strategy is to expand the marketing zone with a higher marketing cost while sacrificing the promotion accuracy, then $\alpha$ needs to be set to a certain large positive value. In our experiments, to cover various situations, it is assumed that $\alpha\in\{0,0.5,1\}$. We assume $\delta=10^{-4}$ for all POIs, which can be set to another value to adjust the convergence speed.

%%%%%%%%%%%%%%%%%%%%%%%%%%%%%%%%%%%%%%%%%%%%%%%%%%%%%%%%%%%
\begin{table}[t]
\centering

\caption{Statistics of the four POIs.}
\label{tab:input}
\begin{tabular}{|p{2.3cm}|l|l|l|p{2.2cm}|}
\hline
\multicolumn{1}{|c|}{\textbf{POI name}}    & $|\mathcal{D}(c^{(0)},\bar{r})|$ &$|\mathcal{D}_{\texttt{all}}(c^{(0)},\bar{r})|$& \begin{tabular}[c]{@{}l@{}}$r_\texttt{cover}$  (m) \end{tabular}\\\hline
Empire State Building                                                           &  1,061\begin{tabular}[c]{@{}l@{}} \end{tabular}&  6,812   & 201\\ \hline
\begin{tabular}[c]{@{}l@{}} Dodger Stadium    \end{tabular}      & 941 &  2,228  &150\\ \hline
Metropolitan  Museum of  Art  &591&     18,413 &239\\ \hline
\begin{tabular}[c]{@{}l@{}}Busch Gardens\end{tabular}  &166&297 & 660\\ \hline
\end{tabular}
\end{table}

\subsection{Estimation Performance}\label{beqisb}

\begin{figure}[t!]
\centering
        \begin{subfigure}[]{0.49\textwidth}
		\begin{subfigure}[]{0.49\textwidth}
		 \centering
            \includegraphics[height=\textwidth, width=\textwidth]{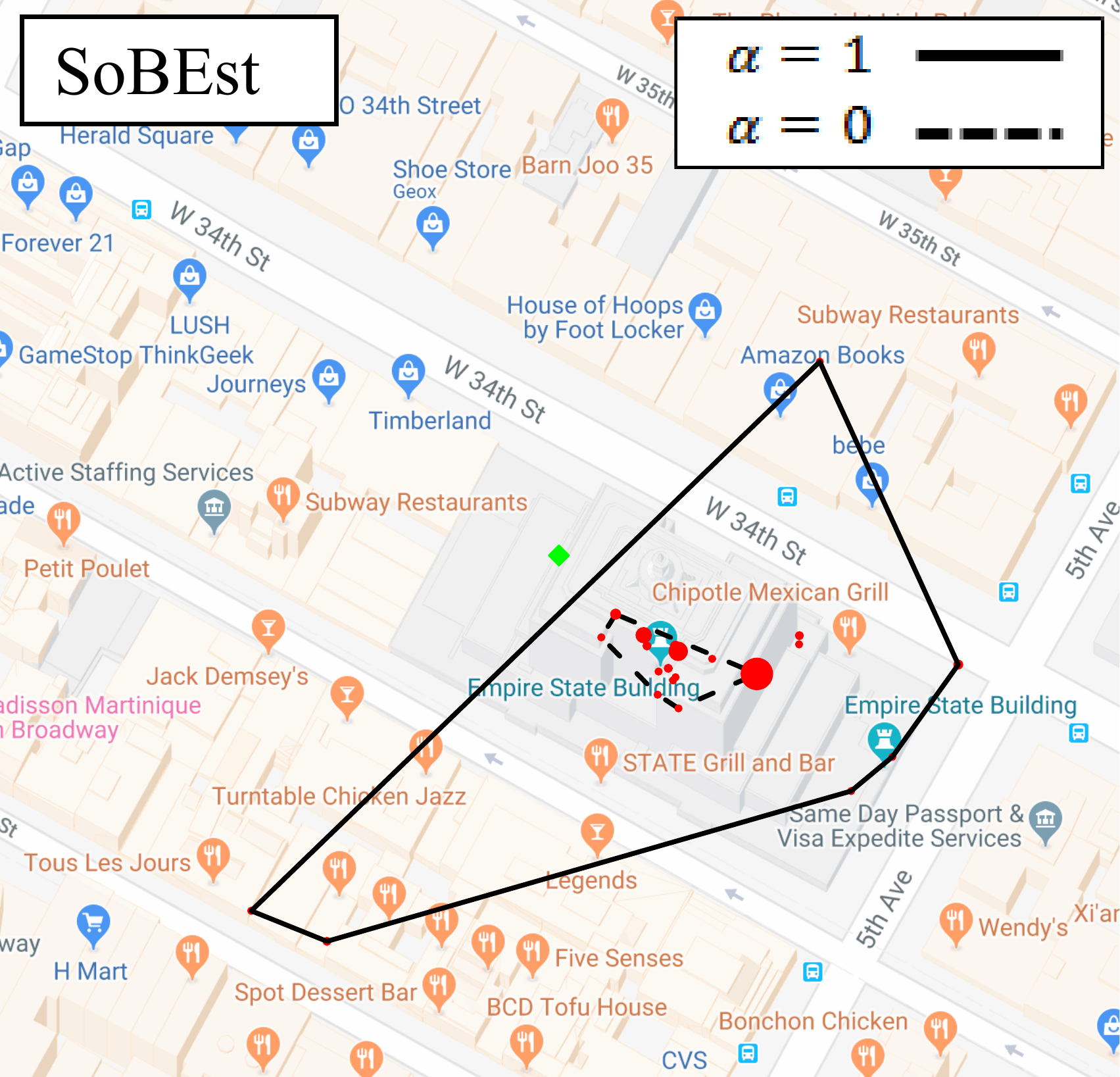}
		\end{subfigure}
		\begin{subfigure}[]{0.49\textwidth}
		 \centering
            \includegraphics[height=\textwidth, width=\textwidth]{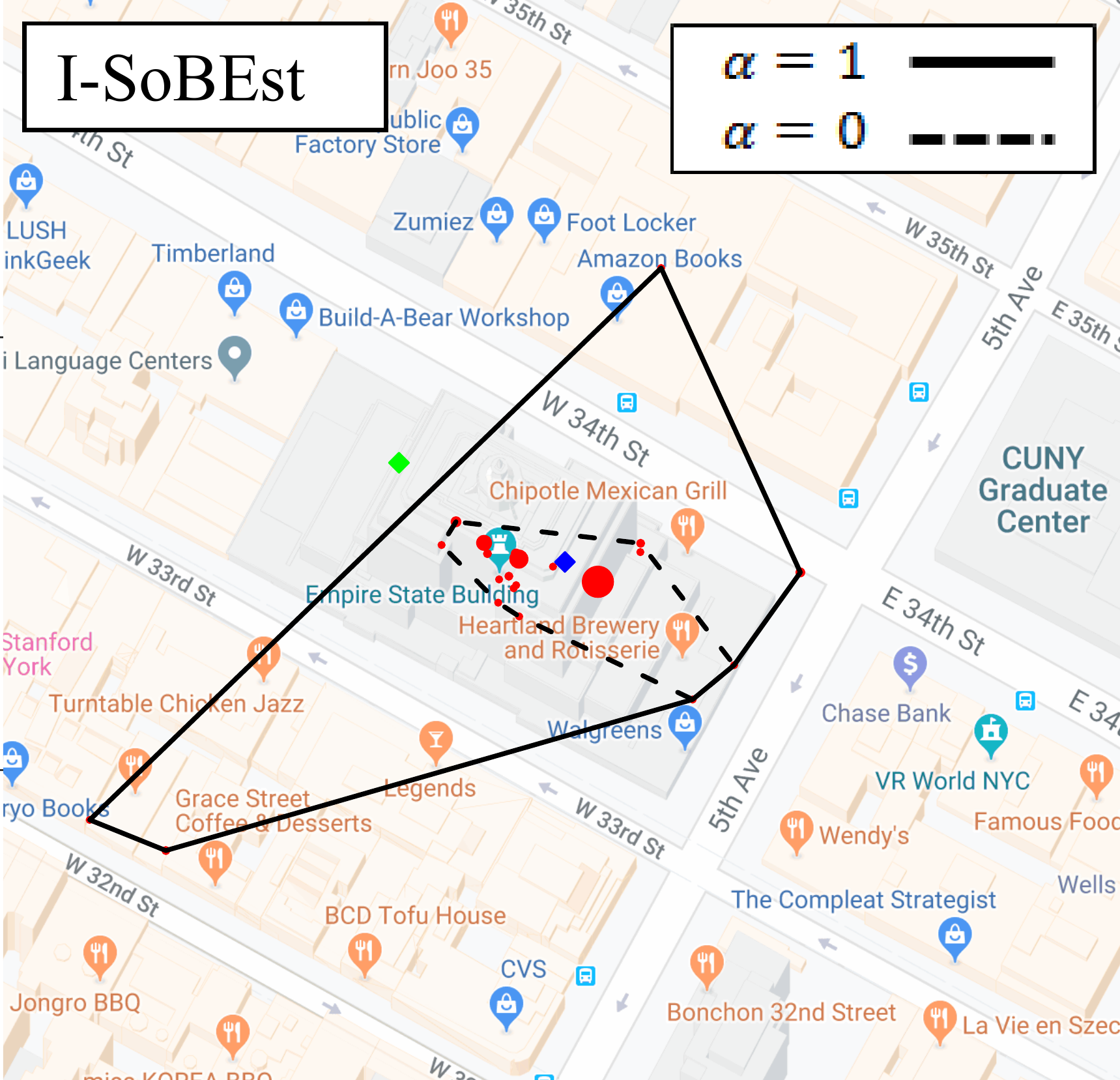}
		\end{subfigure}
		\caption{Empire State Building}
            \label{fig:es}
        \end{subfigure}
        \begin{subfigure}[]{0.49\textwidth}
		\begin{subfigure}[]{0.49\textwidth}
		 \centering
            \includegraphics[height=\textwidth, width=\textwidth]{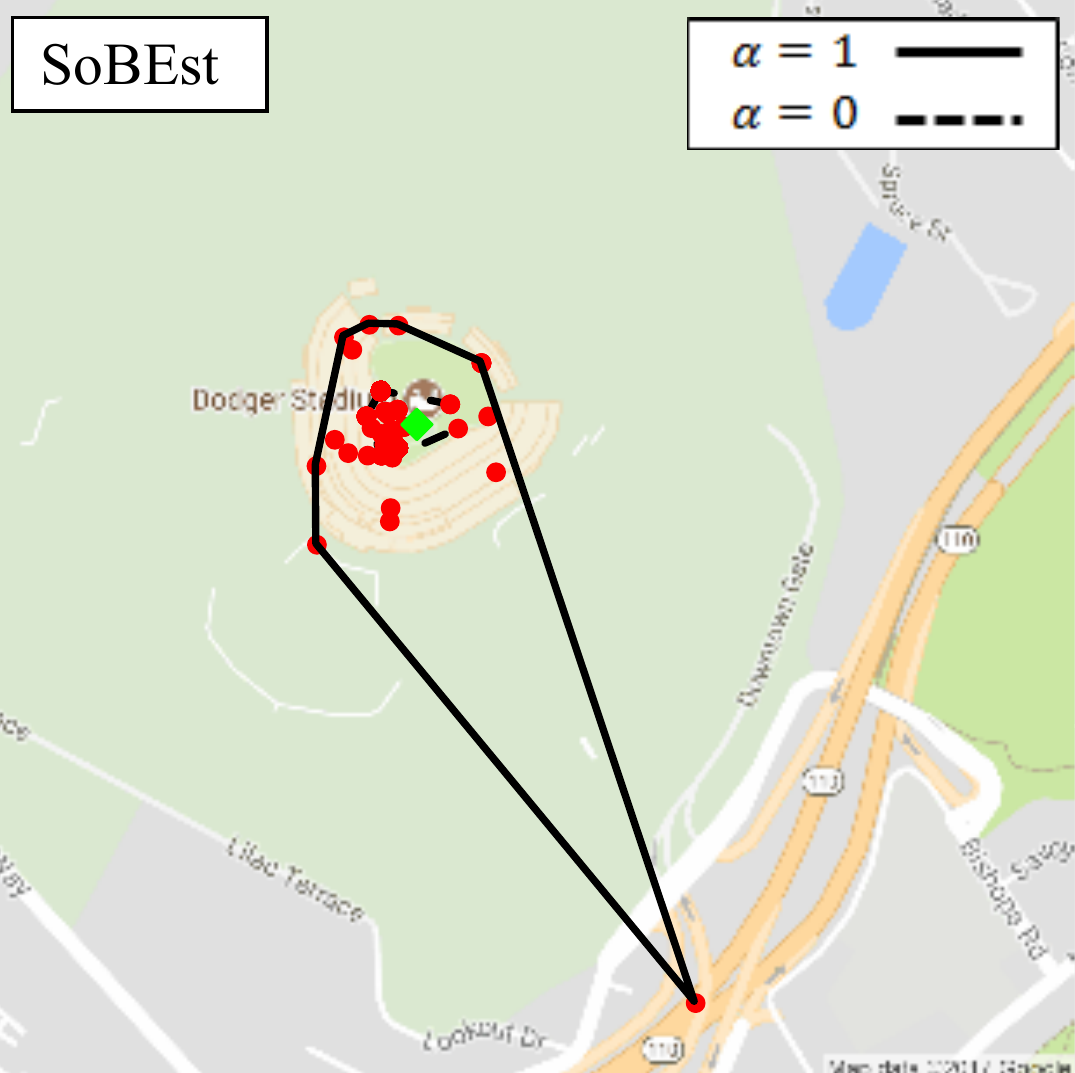}
		\end{subfigure}
		\begin{subfigure}[]{0.49\textwidth}
		 \centering
            \includegraphics[height=\textwidth, width=\textwidth]{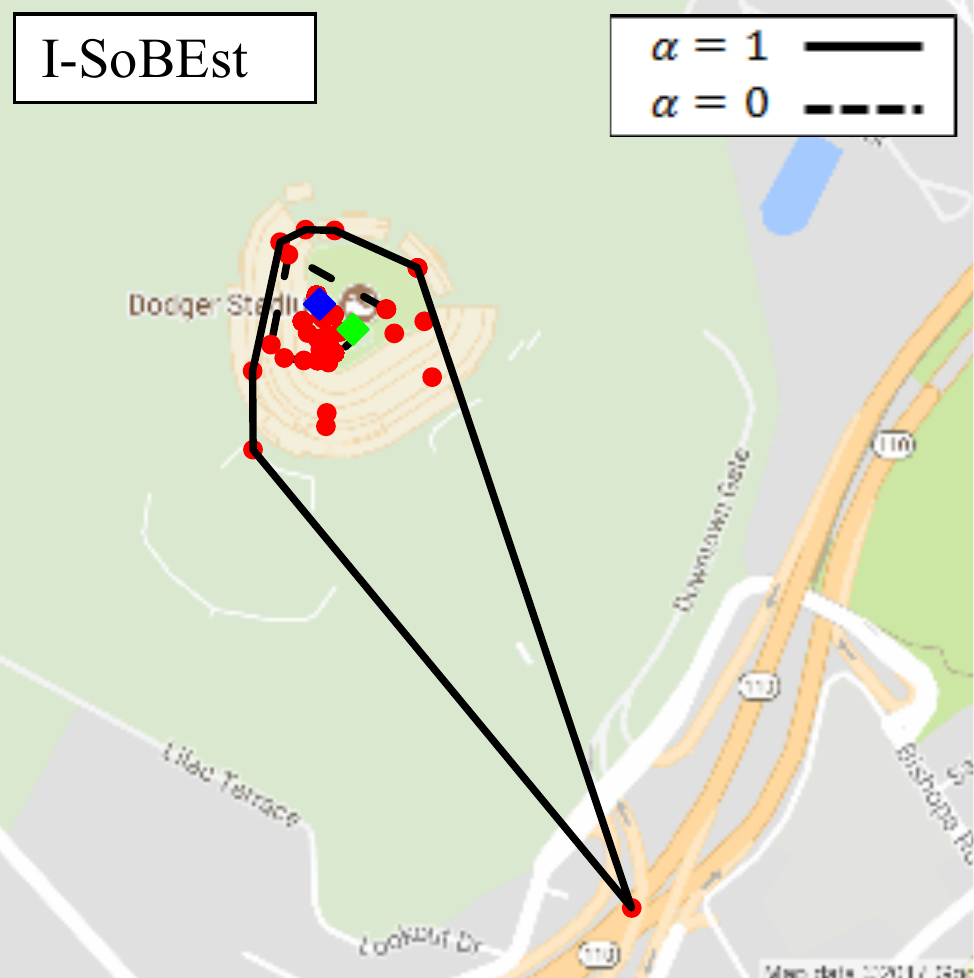}
		\end{subfigure}
            \caption{Dodger Stadium}
            \label{fig:ds}
        \end{subfigure}
        \begin{subfigure}[]{0.49\textwidth}
		\begin{subfigure}[]{0.49\textwidth}
		 \centering
            \includegraphics[height=\textwidth, width=\textwidth]{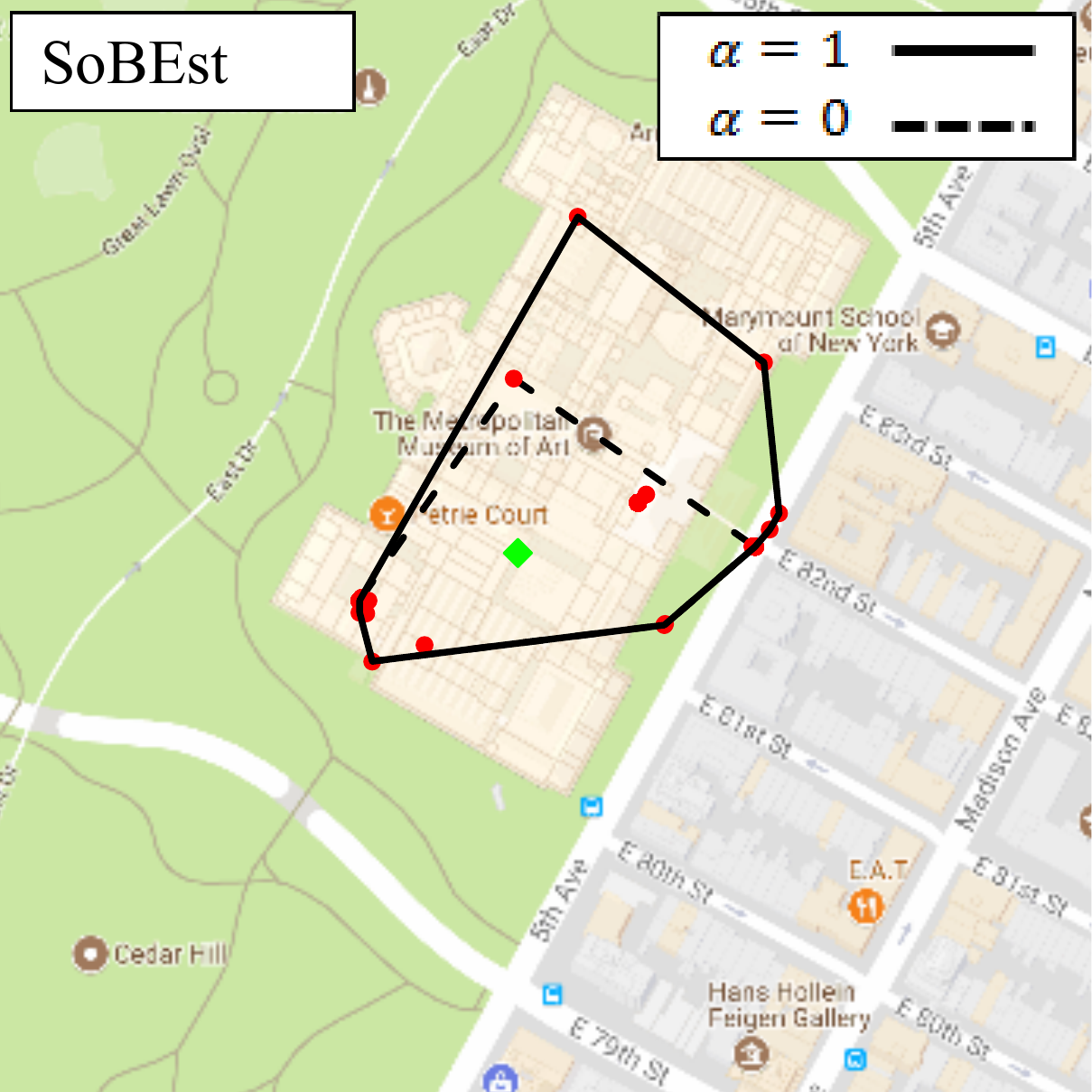}
		\end{subfigure}
		\begin{subfigure}[]{0.49\textwidth}
		 \centering
            \includegraphics[height=\textwidth, width=\textwidth]{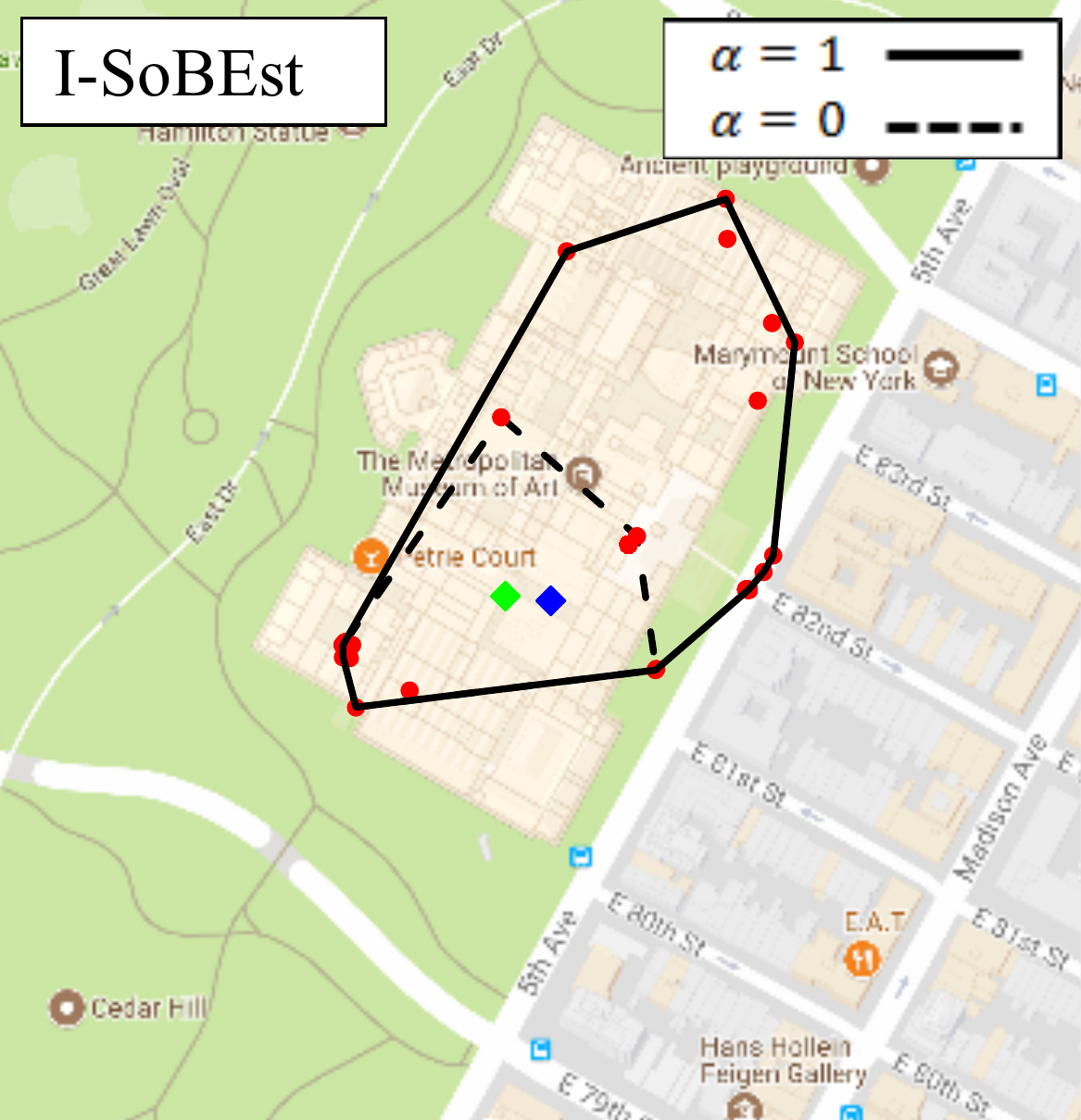}
		\end{subfigure}
		\centering
            \caption{{Metropolitan Museum of Art}}
            \label{fig:tm}
        \end{subfigure}
        \begin{subfigure}[]{0.49\textwidth}
		\begin{subfigure}[]{0.49\textwidth}
		 \centering
            \includegraphics[height=\textwidth, width=\textwidth]{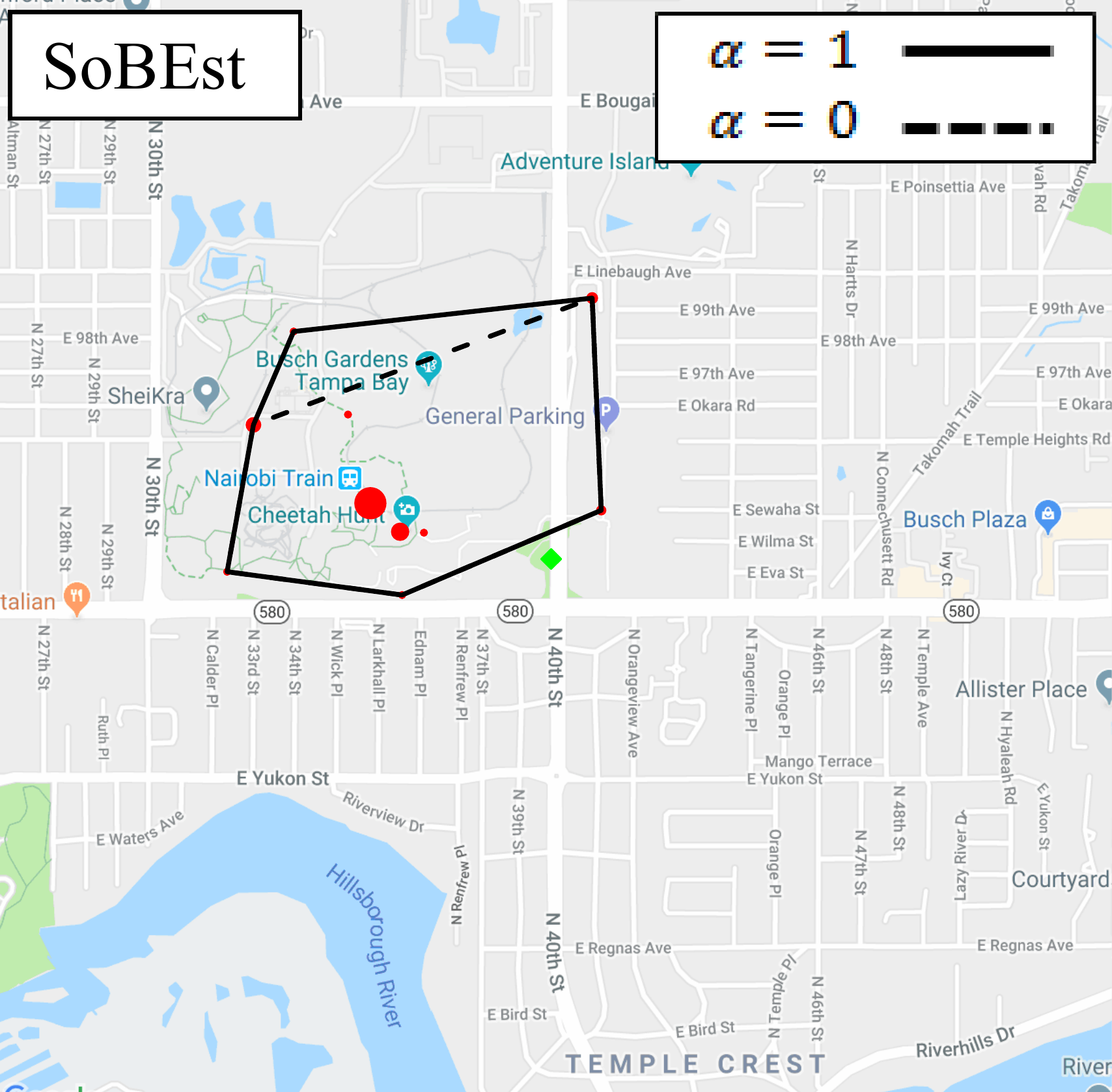}
		\end{subfigure}
		\begin{subfigure}[]{0.49\textwidth}
		 \centering
            \includegraphics[height=\textwidth, width=\textwidth]{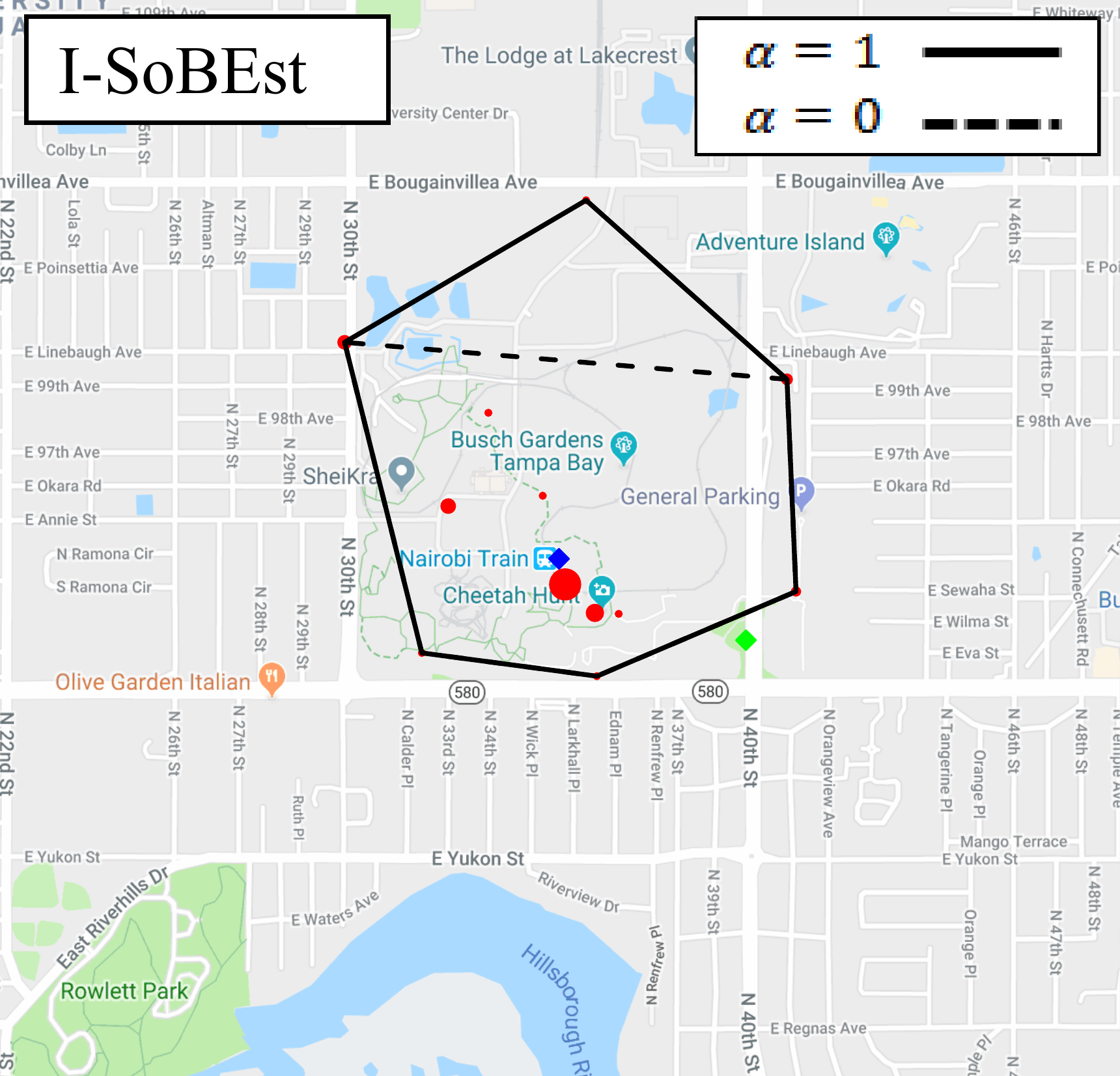}
		\end{subfigure}
            \caption{{Busch Gardens}}
            \label{fig:bg}
        \end{subfigure}
\caption{Social POI boundaries estimated by \textsf{SoBEst} and \textsf{I-SoBEst}, where one green diamond, one blue diamond, and red circles indicate the initial representative coordinate, updated representative coordinate, and relevant records, respectively, in each subfigure. Here, the size of each red circle is increased according to the number of superimposed records.}
\label{fig:al3update}
\end{figure}

Estimation results of the \textsf{I-SoBEst} algorithm are shown according to different values of $\alpha \ge 0$ for the four POIs. When $\delta = 10^{-4}$, the BEQ $\left(\frac{r}{\bar{r}}\right)^\alpha\mathcal{F}(c,r)$ converges to $\left(\frac{r^*}{\bar{r}}\right)^{\alpha}\mathcal{F}(c^*,r^*)$ within three iterations. The BEQ of \textsf{I-SoBEst} for the four POIs is summarized in Table~\ref{tab:estimation2} in comparison with that of \textsf{SoBEst} without updating the POI's representative coordinate $c$, where $\alpha \in \{0,0.5,1\}$. For better understanding, the social POI boundaries estimated by both \textsf{SoBEst} \cite{cGeorich} and \textsf{I-SoBEst} are illustrated in Fig.~\ref{fig:al3update}, where $\alpha \in \{0, 1\}$; the red circles are the relevant records; the green diamond is the initial representative coordinate of each POI; and the blue diamond is the updated representative coordinate of each POI. From Table~\ref{tab:estimation2} and Fig.~\ref{fig:al3update}, the following interesting findings are observed. 
\begin{itemize}
\item When the updated representative coordinate $c^*$ of a POI is far away from the initial one $c^{(0)}$ (e.g., Empire State Building and Busch Gardens), the BEQ of \textsf{I-SoBEst} is enhanced up to $16.75\%$, compared to the original \textsf{SoBEst} case, where the performance improvement over \textsf{SoBEst} is expressed as $\frac{\text{(BEQ of \textsf{I-SoBEst}) - (BEQ of \textsf{SoBEst})}}{\text{BEQ of \textsf{SoBEst}}}\times 100\%$.

\item On the other hand, when the two representative coordinates of a POI marked with green and blue diamonds are close to each other (e.g., Dodger Stadium), the performance improvement is marginal or does not exist.
\end{itemize}

In consequence, the distribution of relevant records (leading to a different centroid) significantly affects the improvement rate of the BEQ. For example, as illustrated in Fig.~\ref{fig:bg}, the social POI boundary of Busch Gardens estimated by \textsf{I-SoBEst} covers a part of the SheiKra roller coaster, which is located in the western part of Busch Gardens, for all $\alpha$'s by iteratively updating the POI's representative coordinate; however, such a tendency is not exhibited for the case estimated by the original \textsf{SoBEst}. Thus, the social POI boundary estimated by \textsf{I-SoBEst} is more appropriate for geomarketing in some cases where a fixed representative coordinate of each POI is far away from the locations in which relevant records were posted.

\begin{table}[t!]
    \centering
    \caption{The BEQ of \textsf{SoBEst} and \textsf{I-SoBEst}.}
    \label{tab:estimation2}
    \begin{tabular}{|p{1.6cm}|l|c|c|}
        \hline 
        \begin{tabular}[c]{@{}l@{}}{\textbf{POI} \textbf{name} }\end{tabular} &{ $\alpha$} &
\begin{tabular}[c]{@{}l@{}}{BEQ of \textsf{SoBEst}}\end{tabular}&
\begin{tabular}[c]{@{}l@{}}{BEQ of \textsf{I-SoBEst}}\end{tabular}  \\ \hline
        \begin{tabular}[c]{@{}l@{}} Empire\\State\\Building  \end{tabular} & \begin{tabular}[c]{@{}l@{}}\begin{tabular}[c]{@{}l@{}}{0}\\0.5\\{1}\end{tabular}\end{tabular}  &\begin{tabular}[c]{@{}l@{}}0.837\\0.187\\0.049\end{tabular}& \begin{tabular}[c]{@{}l@{}}{0.874} \\0.205\\{0.056}\end{tabular} \\ \hline
        \begin{tabular}[c]{@{}l@{}} Dodger\\Stadium   \end{tabular}  & \begin{tabular}[c]{@{}l@{}}\begin{tabular}[c]{@{}l@{}}{0}\\0.5\\{1}\end{tabular}\end{tabular} &\begin{tabular}[c]{@{}l@{}}0.759\\0.566\\0.480\end{tabular}& \begin{tabular}[c]{@{}l@{}}{0.760} \\0.566\\{0.480}\end{tabular} \\ \hline
        \begin{tabular}[c]{@{}l@{}}Metropolitan \\Museum\\of Art\end{tabular}      & \begin{tabular}[c]{@{}l@{}}\begin{tabular}[c]{@{}l@{}}{0}\\0.5\\{1}\end{tabular}\end{tabular}  &\begin{tabular}[c]{@{}l@{}}0.844\\0.224\\0.060\end{tabular}& \begin{tabular}[c]{@{}l@{}}{0.847}\\ 0.236\\{0.066}\end{tabular} \\ \hline
        \begin{tabular}[c]{@{}l@{}}Busch\\Gardens\end{tabular}   & \begin{tabular}[c]{@{}l@{}}\begin{tabular}[c]{@{}l@{}}{0}\\0.5\\{1}\end{tabular}\end{tabular} &\begin{tabular}[c]{@{}l@{}}0.816\\0.304\\0.117\end{tabular}& \begin{tabular}[c]{@{}l@{}}{0.838}\\0.323\\ {0.137}\end{tabular} \\ \hline

    \end{tabular}
\end{table}

\subsubsection{Comparison With Competing Methods}

Now, to verify the superiority of the proposed \textsf{I-SoBEst}  algorithm, let us turn to the performance comparison with two baseline methods and three state-of-the-art methods. In the following, we briefly describe the basic mechanism, parameter settings, and BEQs of each method.

\textbf{DBSCAN  (Baseline \#1)}. As mentioned in the prior studies, DBSCAN is the most commonly used density-based clustering algorithm in identifying the AOI, and thus can be considered as a baseline method in this field of study. Rather than adopting more recently developed state-of-the-art DBSCAN methods in \cite{stdbscan, hdbscan, c15, c11, c10}, we use our own variant by modifying the original DBSCAN algorithm in \cite{c16}. This is because all the DBSCAN methods do not exploit any property of textual attributes in discovering clusters. More precisely, the original one does not take into account any textual attributes and performs clustering based purely on the spatial density. Our modified DBSCAN method is suitable for performance evaluation under our problem setting since it enables us to surely guarantee higher performance. Thus, instead of $\mathcal{D}_\texttt{all}(c^{(0)},\bar{r})$, the set $\mathcal{D}(c^{(0)},\bar{r})$ including solely \textit{relevant records} is assumed to be used as input of DBSCAN, which will result in a much higher BEQ. DBSCAN operates based on two important parameters $MinPts$ and $\epsilon$, where $MinPts$ is the minimum number of points required to form a dense region  within a distance threshold  $\epsilon$ \cite{c16}. Setting $\epsilon$ too high (e.g., $\epsilon$ = 10 km) may result in over-expanded clusters, while setting it too low (e.g., $\epsilon$ = 10 m) may lead to missing large clusters. Our modified DBSCAN method is suitable for performance evaluation under our problem setting since it enables us to guarantee higher performance. Thus, $\epsilon$ needs to be carefully chosen according to each POI type. Moreover, $MinPts$ should be chosen as at least $3$ since otherwise, the result will be the same as the hierarchical clustering case  with the single  link metric \cite{optic}. For a fair comparison with the proposed approach, we then select the one having the largest number of relevant records among multiple clusters, named  the \textit{POI cluster}, as a counterpart of the social POI boundary estimated from \textsf{SoBEst}. Thereafter, using (\ref{eq:1}) and (\ref{eq:2}), the $\mathcal{F}$-measure of a given POI cluster can be computed by replacing $\mathcal{D}_\texttt{all}(c^{(0)},r)$ and $\mathcal{D}(c^{(0)},r)$ by the set of all geo-tags and the set of all relevant records, respectively, only within the POI cluster. In our work, we assume $MinPts = 5$ and $\epsilon= r_\texttt{cover}$, which are properly chosen to return a high $\mathcal{F}$-measure for the selected POI cluster.

\textbf{One-class support vector machine (OCSVM)  (Baseline \#2)}. OCSVM \cite{svm} is a well-known unsupervised clustering technique for learning a decision boundary between  outliers and the rest of a given  dataset, in which an input parameter $\vartheta$ specifies the probability of outliers. In \cite{ocsvm}, the OCSVM algorithm was employed to estimate a crisp boundary of an AOI---the set of relevant geo-tagged records, corresponding to $\mathcal{D}(c^{(0)},\bar{r})$, was exploited as input of the algorithm, where the parameter $\vartheta$ is set to 0.14 based on the statistical analysis on regions  with administrative boundaries. Since this parameter setting results in a high $\mathcal{F}$-measure for each  selected POI cluster in our experiments, it is adopted to evaluate the BEQ of OCSVM.

\textbf{Two-phase estimation for POI boundaries (2P-BEst)  (State-of-the-art \#1)}. A low-complexity two-phase POI boundary estimation (2P-BEst) algorithm was developed in \cite{c23} as a state-of-the-art method  by discovering a circle with radius $r^*$ that covers one POI cluster. The algorithm takes $\mathcal{D}(c^{(0)},\bar{r})$,  which is the set of relevant records within a circle $(c^{(0)},\bar{r})$, as the input along with the following three tuning parameters: the radius interval $\Delta r_1$  for the first phase, the interval granularity $Q$, and the threshold $\Delta\eta$  determining the update condition for the first phase. In our experiments, $\Delta\eta$ and $Q$ are set to 100 and 10, respectively, as in \cite{c23}, while $\Delta r_1$ is set to $r_\texttt{cover}$, which is determined adaptively according to a given POI and thus yields a relatively high $\mathcal{F}$-measure.

\textbf{Fast density peak clustering (FastDPeak)  (State-of-the-art \#2)}. We implement an efficient DPC algorithm, dubbed FastDPeak~\cite{newref4}, as another state-of-the-art method that detects the largest POI cluster based on the density of relevant records (i.e., the set $\mathcal{D}(c^{(0)},\bar{r})$). FastDPeak tends to perform quite poorly for four selected POIs due to the existence of superimposed records since the local density of such records becomes extremely high. Thus, the original algorithm is modified in such a way that 90\% of superimposed records are empirically removed to achieve a high BEQ through intensive experiments, which can be well generalized over all four selected POIs. In our experiments, an input parameter $C$, indicating the number of detected clusters, is properly chosen to return a high BEQ for the selected POI cluster.

\textbf{Robust noise resistent POI identification (RNRPI)  (State-of-the-art \#3)}. We employ another state-of-the-art  algorithm of detecting boundaries of multiple POIs in \cite{lzpd}. This algorithm introduces a technique incorporating Laplacian zero-crossing points, which adaptively filters noise records from the set $\mathcal{D}(c^{(0)},\bar{r})$ based on the local drastic changes of the data density given by a Gaussian kernel density estimation function. Since the density estimation function cannot perform properly due to the presence of superimposed records, we also remove 90\% of superimposed records as in FastDPeak to achieve a higher degree of the BEQ for  four selected POIs. In our experiments, the parameter $\alpha$ corresponding to the step size for the gradient of the density estimation function in RNRPI is set appropriately for each POI.

\begin{figure}[t]
\centering

        \begin{subfigure}[]{\linewidth}
            \includegraphics[width=\linewidth]{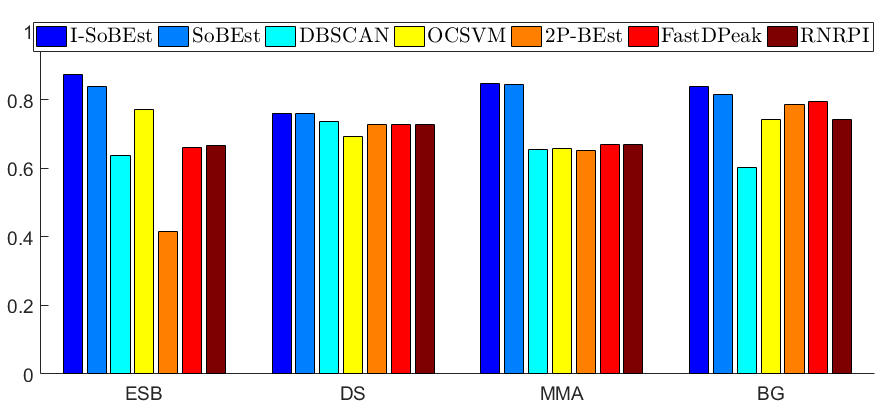}
            \centering
            \caption{ $\alpha = 0$}
            \label{fig:110}
        \end{subfigure}
        \begin{subfigure}[b]{\linewidth}

            \includegraphics[width=\linewidth]{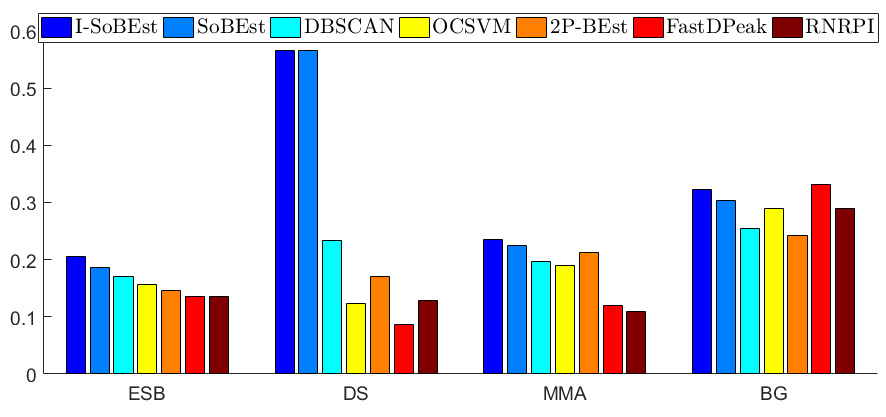}
            \centering
            \caption{ $\alpha = 0.5$ }
            \label{fig:210}
        \end{subfigure}
        \begin{subfigure}[b]{\linewidth}

            \includegraphics[width=\linewidth]{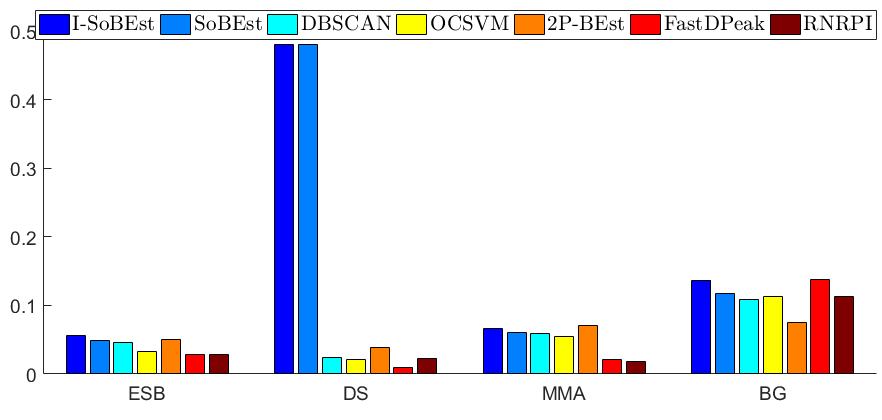}
            \centering
            \caption{ $\alpha = 1$}
            \label{fig:111}
        \end{subfigure}
        \caption{The performance comparison of \textsf{I-SoBEst},  \textsf{SoBEst}, DBSCAN, OCSVM, 2P-BEst, FastDPeak, and RNRPI algorithms, where the BEQ is illustrated according to each POI, and ESB, DS, MMA, and BG indicate Empire State Building, Dodger Stadium, Metropolitan Museum of Art, and Busch Gardens, respectively.}
        \label{fig:comp}
\end{figure}

\textbf{Performance comparison.} Due to the fact that four algorithms, including DBSCAN, OCSVM, FastDpeak, and RNRPI return arbitrarily-shaped clusters, we need to find a proper radius $r_\texttt{D}$ of the POI cluster  in order to evaluate the BEQ of these methods under our problem setting. In our work, the radius $r_\texttt{D}$ is given by the maximum distance between all relevant records in the cluster and the centroid of all the relevant records in the cluster. Finally, we present performance comparison among \textsf{SoBEst} and \textsf{I-SoBEst} as well as five competing methods in terms of BEQ in Fig.~\ref{fig:comp}. From the figure, the following insightful observations are found. 
\begin{itemize}
\item \textsf{I-SoBEst} outperforms all other methods in terms of BEQ by up to an order of maginitude in almost all cases (refer to the Dodger Stadium for $\alpha=1$ to see the largest improvement over FastDPeak).
\item When the initial representative coordinate  $c^{(0)}$ is not correctly placed, \textsf{I-SoBEst} can alleviate the weakness of \textsf{SoBEst} (refer to the Busch Garden for $\alpha=1$ to see additional performance gain). The effect of the initial representative coordinate on the BEQ can also be identified by comparing the results of \textsf{SoBEst} and \textsf{I-SoBEst} algorithms in Figs.~\ref{fig:al3update} and~\ref{fig:comp}.
\end{itemize}

\subsection{Computational Complexity}
\begin{figure}[t]
\includegraphics[width=\columnwidth]{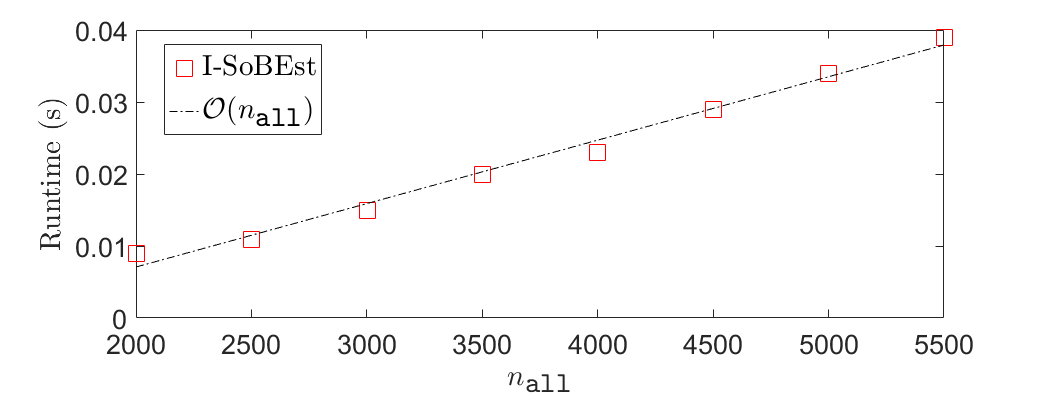}
\centering
\caption{Runtime complexity for estimating the social POI boundary of the Empire State Building.}
\label{fig:complexity}
\end{figure}

In this subsection, we analyze the computational complexity of \textsf{I-SoBEst} (Algorithm 3). Let us denote the cardinality of the set $\mathcal{D}_\texttt{all}(c^{(0)},\bar{r})$ by $n_\texttt{all}\triangleq |\mathcal{D}_\texttt{all}(c^{(0)},\bar{r})|$, representing the number of all geo-tagged tweets in a circle $(c^{(0)},\bar{r})$.
In our \textsf{I-SoBEst} algorithm, the overall computational complexity is dominated by \textsf{SoBEst} whose runtime complexity is given by $\mathcal{O}(n_\texttt{all})$ according to the argument in \cite{cGeorich}. More specifically, it is shown that the computational complexity still scales as $\mathcal{O}(n_\texttt{all})$ (i.e., linearity in $n_\texttt{all}$) as long as the number of iterations is very small and, more importantly, is {\em independent} of $n_\texttt{all}$. We numerically validate this argument by examining that the number of iterations is {\em at most three} when $\delta = 10^{-4}$ and the POIs in Table~\ref{tab:input} are used. Next, in Fig.~\ref{fig:complexity}, the overall runtime is evaluated when $n_\texttt{all}$ varies from 2,000 to 5,500 in an increment of 500, where the social POI bound of Empire State Building is estimated for $\alpha=1$ and $n_\texttt{all}$ points are randomly sampled from 6,812 geo-tagged records for the Empire State Building. An asymptotic dotted line with a proper bias is also plotted in Fig.~\ref{fig:complexity}, showing trends consistent with our experimental result.

%%%%%%%%%%%%%%%%%%%%%%%%%%%%%%%%%%%%%%%%%%%%%%%%%%%%%%%%%%%%

\section{Concluding Remarks}
\label{sec:conclusion}

In this paper, we introduced an improved framework for estimating social POI boundaries, which has a broad range of applications from geomarketing, event management to urban planning, along with an iterative estimation method. Based on our prior estimation approach exploiting the spatio--textual data on social media, we presented a new joint optimization of the radius of a circle and the POI's representative coordinate $c$ by allowing to update $c$, and then proposed the \textsf{I-SoBEst} algorithm. To show the effectiveness of our algorithm, we numerically evaluated the BEQ for various $\alpha$'s and POIs, and delineated the estimated social POI boundaries. It turned out that the \textsf{I-SoBEst} algorithm is superior to the original \textsf{SoBEst} by up to 16.75\% and remarkably outperforms the baseline and state-of-the-art methods by up to an order of magnitude for almost all cases. In addition, the computational complexity of \textsf{I-SoBEst} was analytically shown and numerically validated. This sheds light on the design of density-based clustering by intelligently integrating spatial and textual attributes on social media.

\section*{Acknowledgments}

This research was supported by the Republic of Korea’s MSIT (Ministry of Science and ICT), under the High-Potential Individuals Global Training Program (No. 2020-0-01463) supervised by the IITP (Institute of Information and Communications Technology Planning Evaluation), by a grant of the Korea Health Technology R\&D Project through the Korea Health Industry Development Institute (KHIDI), funded by the Ministry of Health \& Welfare, Republic of Korea (HI20C0127), and by the Yonsei University Research Fund of 2020 (2020-22-0101). This paper was presented in part at the ACM SIGMOD Workshop on Managing and Mining Enriched Geo-Spatio Data, San Francisco, CA, June 2016 \cite{cGeorich}. 

% Bibliography
\bibliographystyle{cas-model2-names}
\bibliography{Elsevier_KBS}

\end{document}